\documentclass[reprint,prb,amsmath,amssymb,superscriptaddress]{revtex4-2}
\usepackage{xcolor}
\usepackage{graphicx}% Include figure files
\usepackage{bm}% bold math
\usepackage[colorlinks,citecolor=blue,linkcolor=red]{hyperref}

\def\be{\begin{equation}}
\def\ee{\end{equation}}
\def\bd{\begin{displaymath}}
\def\ed{\end{displaymath}}
\def\-{\phantom{-}}

\begin{document}

\title{Parameter free treatment of a layered correlated van der Waals magnet: CrPS$_{4}$}

\author{A. R. Alcantara}
\affiliation{Department of Physics, University of North Florida, Jacksonville, FL 32224}
\affiliation{Theoretical Division, Los Alamos National Laboratory, Los Alamos, NM, 87545}
\author{C. Lane}
\affiliation{Theoretical Division, Los Alamos National Laboratory, Los Alamos, NM, 87545}
\author{J. T. Haraldsen}
\affiliation{Department of Physics, University of North Florida, Jacksonville, FL 32224}
\author{R. M. Tutchton}
\affiliation{Theoretical Division, Los Alamos National Laboratory, Los Alamos, NM, 87545}
\date{\today}

\begin{abstract}
The electronic and magnetic structure of CrPS$_{4}$, a 2D magnetic semiconductor is examined by employing the SCAN meta-GGA density functional. We find the resulting magnetic moment and band gap are in excellent agreement with experiment. From the bulk magnetic configurations, we confirm the experimentally observed A-type antiferromagnetic (A-AFM) ordered ground state with a magnetic moment of 2.78 $\mu_B$ per chromium atom and band gap of 1.34 eV. To gain insight into the evolution of the ground state with layers, the total energy of each magnetic configuration is calculated for a variety of thicknesses. Monolayer CrPS$_{4}$ is predicted to be a ferromagnetic insulator with a band gap of 1.37 eV, and A-AFM for bilayer and trilayer, with band gaps of 1.35 eV and 1.30 eV, respectively. The electronic structure is reported for the single, two, three layer and bulk CrPS$_{4}$.  Finally, we explore the optical properties of the 2D structure and report the dielectric tensor components and Kerr parameters for the monolayer.
\end{abstract}

\maketitle

\section{Introduction}
Recently, several reports of new 2D magnetic materials have appeared in the literature displaying a rich variety of magnetic structures~\cite{gibe:19}, along with a myriad of competing topological and superconducting phases~\cite{li2021electronic,kim2018large}. The layer-dependent ferromagnetism exhibited by these compounds, such as CrI$_3$, offers a new pathway for the development of new spintronic devices because of their wide tunabilities using doping, electric field, light, and pressure~\cite{burch2018magnetism}. However, the current workhorse CrI$_3$, and similar materials, are fundamentally limited due to their extreme sensitivity to air~\cite{shcherbakov2018raman}, requiring special glovebox environments and capping layers to preserve the material properties. Therefore, it is crucial to identify air-stable 2D ferromagnetic materials to advance the next generation spintronic, optoelectronic, and future quantum information technologies in general~\cite{gibe:19}.

CrPS$_4$ has recently been singled out due to its intrinsic ferromagnetic ordering in the monolayer~\cite{cald:20} and robust air-stability~\cite{sonj:21} similar to CrTe$_2$ \cite{meng2021anomalous,purbawati2023stability,huey2022cr}. These key proprieties have prompted a number of experimental~\cite{peiq:16,lee:17,budn:20,gupi:19,kim:19,kim:21,zhan:21,shin:21,ries:22,kim:22,xu:22} and theoretical~\cite{zhua:16,joe:17,chen:20,deng:21,yang:21} studies. Specifically, CrPS$_{4}$ is found to exhibit a canted ferromagnetic order within each van der waals layer. When stacked, no net magnetic moment is admitted for an even number of layers characteristic of an A-type AFM ground state~\cite{loui:78}. The large magnetic polarization on each Cr site (2.81 $\mu_B$/Cr) produces a 1.40 eV  gap in the electronic states, suggesting CrPS$_{4}$ is well suited for switching~\cite{peiq:16} and neuromorphic computing~\cite{lee:18} applications. 

The limited theoretical works analyzing CrPS$_4$ have found mixed success. Density functional theory calculations at the generalized gradient approximation (GGA) level are able to find the correct A-AFM magnetic ground state, however, the Cr magnetic moments are underestimated (2.58 $\mu_B$/Cr), along with the band gap (0.79 eV) by almost a factor of two~\cite{zhua:16,loui:78}, as expected for Kohn-Sham theory. To remedy this, a Hubbard $U$ parameter has been applied to the Cr-$3d$ states to yield the correct magnetic moment~\cite{joe:17}. Unfortunately, this correction predicts a X-type AFM ground state in contrast to neutron scattering measurements~\cite{loui:78} and an exaggerated band gap of 1.66 eV. Due to the sensitivity of this system, there is currently no theoretical treatment that captures the delicate balance between the charge and magnetic degrees of freedom. Thus making predictions of magnetoelastic coupling, magneto resistive switching, and magnetic excitations quite challenging.

In this article, we present an accurate parameter-free description of the layer dependent magnetic and electronic properties of CrPS$_{4}$. Utilizing the newly constructed Strongly-Constrained-and-Appropriately-Normed (SCAN) meta-GGA density functional, the calculated magnetic moments (magnitude and direction) and electronic band gap are in excellent accord with reported experimental values. Our predicted magnetic ground state is in agreement with neutron diffraction studies, thereby overcoming the failure of the DFT+$U$ framework. Upon exfoliating the bulk crystal, we find the interlayer coupling to be extremely weak, between 36 to 49 meV, with the corresponding electronic bands exhibiting essentially no change in band dispersion with thickness. Finally, the dielectric tensor and Kerr parameters are obtained for the monolayer to stimulate further experimental works.

\begin{table}
\begin{ruledtabular}
\footnotesize
\caption{Comparison between experimental and various DFT treatments of the magnetic \cite{cald:20} and electronic \cite{loui:78} properties of bulk CrPS$_{4}$. The SCAN results (this work) show improvement over previous PBE \cite{zhua:16} and PBE+U (U=3 eV) \cite{joe:17} studies in capturing the experimental values.}
\begin{center}
\begin{tabular}{ |c|c|c|c|c| }
 Bulk CrPS$_{4}$ &  {\parbox[t]{0.0805\textwidth}{ SCAN}} & {\parbox[t]{0.09\textwidth}{ Experiment}} &  {\parbox[t]{0.075\textwidth}{PBE}} & {\parbox[t]{0.075\textwidth}{ PBE + U}}  \\
  \hline
   Ground state & A-AFM & A-AFM & A-AFM & X-AFM  \\
 \hline
 Band gap (eV) & 1.34 & 1.40 & 0.79 & 1.66   \\
 \hline
Moment ($\mu$B/Cr)  & 3.00 & 2.81 & 2.58 & 3.00  \\

\end{tabular}
\end{center}
\label{prevstudies}
\end{ruledtabular}
\end{table}

%Fig 1 crystal structure
\begin{figure}[h!]
\centering
\includegraphics[width = 0.99\columnwidth]{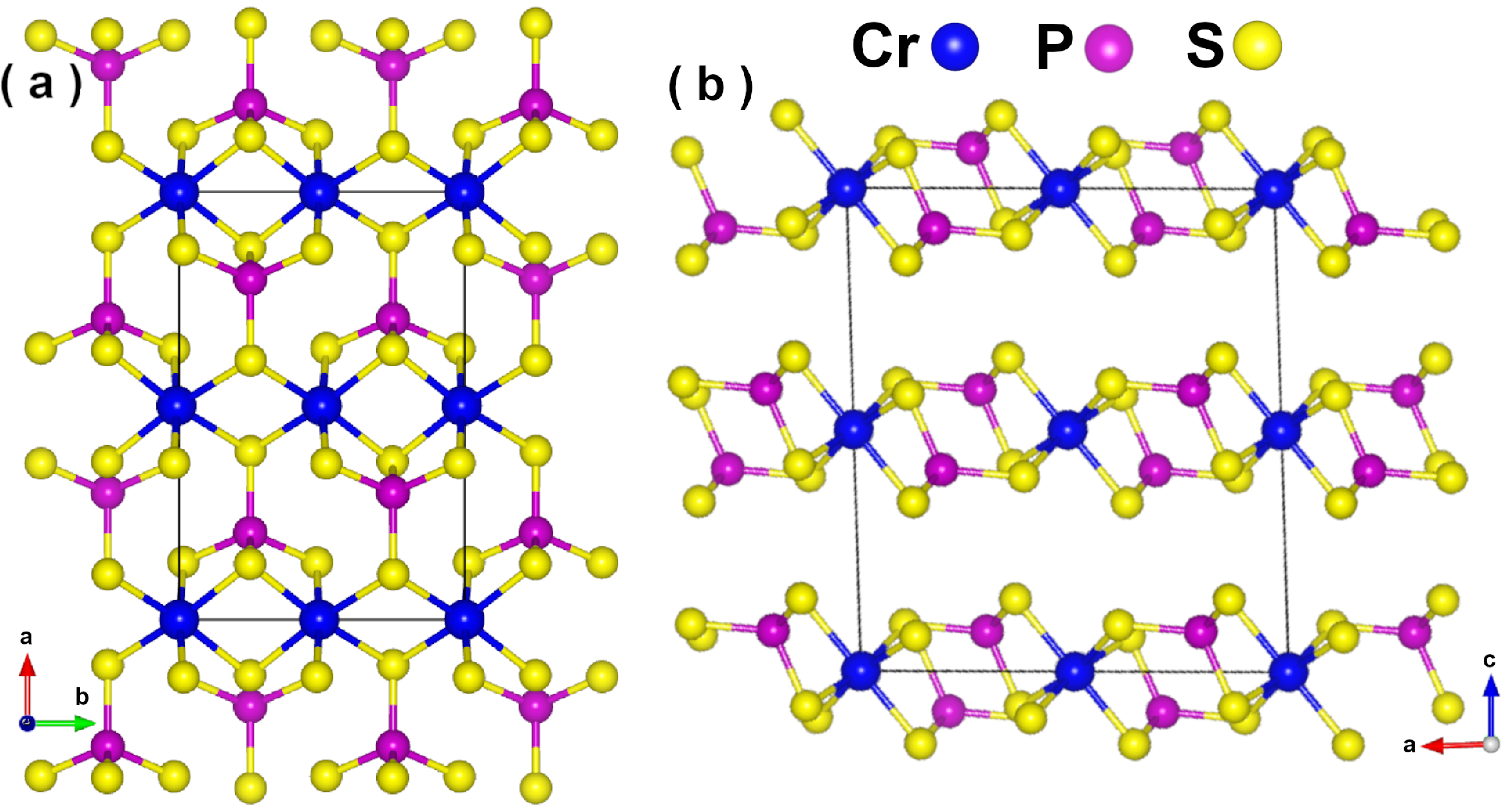}
\caption{Supercell of bulk CrPS$_{4}$ along (a) the $c$-axis and (b) the $b$-axis.}
\label{fig:crystalstructure} 
\end{figure}

%Fig 2 mag configs
\begin{figure*}[ht!]
\centering
\includegraphics[width = 0.99\textwidth]{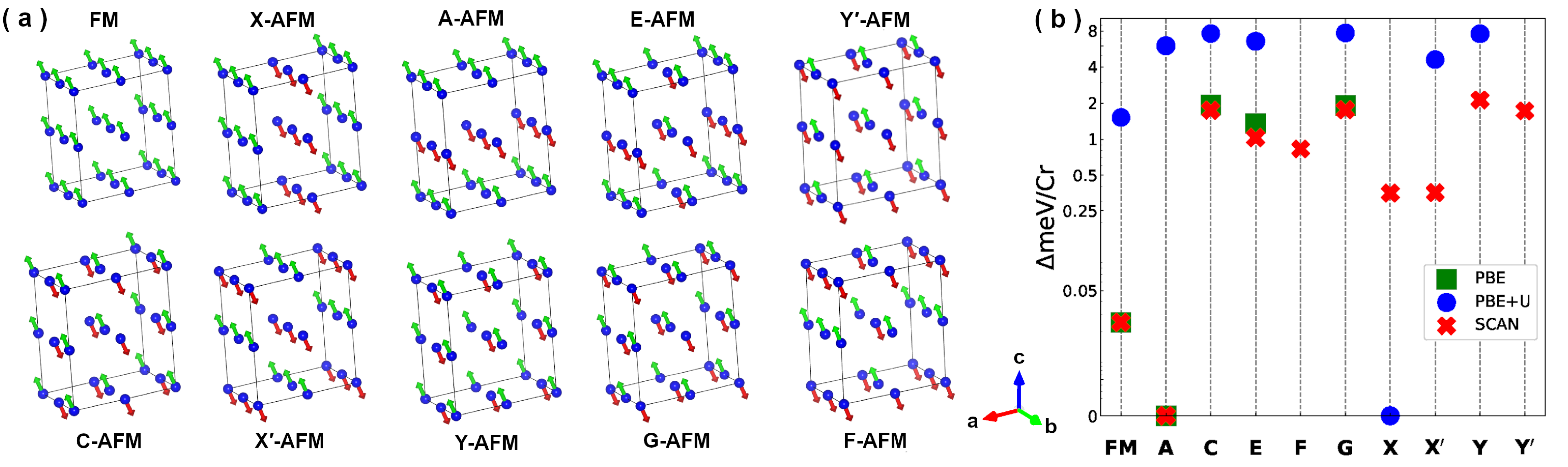}
\caption{(a) Ferromagnetic and the nine antiferromagnetic configurations considered for the ground-state ordering of CrPS$_{4}$. The green and red arrows represent the positive and negative magnetic moments of Cr in the unit cell, respectively. (b) Ground-state energies relative to most stable magnetic configuration by calculated by the PBE (green)\cite{zhua:16}, PBE+U (blue)\cite{joe:17}, and SCAN (red) functionals.}
\label{fig:magconfigs} 
\end{figure*}

{\it Computational Details.}---Calculations were carried out by using the pseudopotential projector-augmented wave (PAW) method~\cite{kres:99} implemented in the Vienna {\it ab initio} simulation package (VASP)~\cite{kres:93,kres:96} with an energy cutoff of 260 eV for the plane-wave basis set. Exchange-correlation effects were treated by using the SCAN, meta-GGA scheme~\cite{sunj:15}. A $6\times7\times1$  $\Gamma$-centered $k$-point mesh was used to sample the single, two, and three layer structures. Whereas a $6\times7\times5$ mesh was used to sample the bulk material. After a systematic study of k-mesh densities, the $6\times7\times5$ mesh was found to converge the total energy to within 0.31 meV/Cr. The dielectric tensor components were calculated using 1296 virtual states. Spin-orbit coupling effects were included self-consistently. The CrPS$_{4}$ unit cell was built using experimentally obtained atomic positions and lattice parameters for the bulk, trilayer, bilayer and monolayer monoclinic symmetry with a space group $C2/m$~\cite{dieh:76}. The magnetism is driven by the Cr$^{3+}$ valence, with three 3$d$ electrons spread in the $t_{2g}$ sub-orbital occupations, yielding a spin moment of $S$ = $3/2$. A vacuum spacing of 19 \AA~ was used to ensure no interactions occur between the periodic images. A total energy tolerance of 10$^{-6}$ eV was used to determine the self-consistent charge density.

\section{Crystal Structure}
Figure~\ref{fig:crystalstructure} shows the crystal structure of bulk CrPS$_{4}$ along the (a) $c$- and (b) $b$-axes. The chromium atoms (blue) sit on an orthorhombic lattice and are octahedrally coordinated by sulfur atoms (yellow) that are slightly distorted off center. Due to the orthorhombic Bravais lattice, the Cr atoms form linear chains of strong inter-site coupling along the $b$-axis, with weaker inter-chain interactions along the $a$-axis. The phosphorus atoms (purple) sit at the center of a sulfur tetrahedral cage, where its position alternates between above and below the Cr layer following the distorted sulfur octahedra. This puckering of the phosphorus sites constrains the primitive unit cell to accommodate four Cr atoms, rather than simply one. Moreover, the phosphorus atoms form a bridge between Cr chains, thereby facilitating the weak inter-chain interactions. Along the $c$-axis, the atomically-thin layers are found to stack in an AA manner with a clear 3.75 \AA~vdW gap separating neighboring layers. This suggests the electronic states should be predominantly 2D in nature with very weak $k_z$ dispersion.

\begin{table}[b]
\caption{ Energy differences ($\Delta$E) relative to the predicted ground states of eight magnetic configurations for the bulk, bilayer, and trilayer, and two magnetic configurations for the monolayer.}
\begin{center}
\begin{tabular}{ |c|c|c|c|c| }
\hline
  {\parbox[t]{0.075\textwidth}{Magnetic\\ order}}  & {\parbox[t]{0.095\textwidth} {Bulk \\ $\Delta$E \\ (meV/Cr)}} &   {\parbox[t]{0.095\textwidth} {Trilayer \\ $\Delta$E \\ (meV/Cr)}}  &  {\parbox[t]{0.095\textwidth}{ Bilayer \\ $\Delta$E \\ (meV/Cr)}} & {\parbox[t]{0.095\textwidth}{ Monolayer $\Delta$E \\ (meV/Cr)}} \\
  \hline
  FM             &  0.031 & 0.075   & 0.088  & 0  \\
   \hline
  A-AFM          &  0       & 0       & 0        & -  \\
 \hline
  C-AFM          &  2.172  & 2.218 & 2.235    & 2.706  \\
\hline
  E-AFM          &  1.288 & 1.312  & 1.286  & -  \\
\hline
  G-AFM          &  2.203 & 2.207  & 2.266  & -  \\
  \hline
  X-AFM          &  0.442  & 2.687  & 2.703  & -  \\
 \hline
  X$^\prime$-AFM &  0.445   & 2.687  & 2.735    & -  \\
\hline
  Y-AFM          &  2.672  & 0.467  & 0.493  & -  \\
\hline
\end{tabular}
\end{center}
\label{toten}
\end{table}

\section{Magnetic Configurations and Ground State}
Figure~\ref{fig:magconfigs} (a)-(b) shows the various collinear commensurate magnetic orders studied in bulk CrPS$_{4}$, along with the corresponding relative total energies of each magnetic configuration with values from previous PBE \cite{zhua:16} and PBE+U (U=3 eV) \cite{joe:17} studies overlayed. Since the bulk primitive cell of CrPS$_4$ has eight Cr atoms spanning two layers, nine distinct commensurate magnetic orders may be accommodated. To enumerate the phases, we initially assume FM coupling between all Cr sites to produce a FM phase with all magnetic moments pointing slightly canted off the $c$-axis. If inter-chain (intra-chain) interactions are switched to AFM, we perturb away from the FM phase yielding a stripe-like X-type (Y-, Y$^\prime$-type) AFM order that varies along the $a$-axis ($b$-axis). Additionally, swapping the FM inter-layer coupling for AFM triggers an A-type AFM state where the direction of the magnetic moments alternate with CrPS$_4$ layers. Finally, by combining inter- and intra-chain, and inter-layer couplings the E-, F-, C-, X$^\prime$-, and G-type AFM states are found.

Figure~\ref{fig:magconfigs} (b) plots the energy of each magnetic configuration relative to the ground state using PBE~\cite{zhua:16} (green squares), PBE+$U$ (U=3 eV)~\cite{joe:17} (blue circles), and SCAN (red $\times$) functionals. PBE correctly captures the correct ground state magnetic order and predicts the FM phase only 2 meV higher in energy. Moreover, the magnetic moment direction and magnitude, though slightly underestimated [Table \ref{prevstudies}], are also in good accord with powder and single crystal magnetic neutron diffraction results \cite{cald:20}. The remaining magnetic arrangements considered (C-, E-, and G-AFM) are separated from the A-AFM phase by $\sim 2$ eV/Cr, making them irrelevant to the low energy degrees of freedom. The close energetic proximity of FM and A-AFM states suggests a delicate balance between AFM and FM coupling between the CrPS$_4$ layers. Despite PBE's success in describing the magnetic state, it severely underestimates the electronic band gap [Table \ref{prevstudies}] as is expected since the Kohn-Sham eigenvalues do not provide the value of the gap. To improve upon the PBE description of the ground state, Joe {\it et al.}~\cite{joe:17} included a Hubbard $U$ of 3 eV on the Cr-$d$ states. By applying a $U$, charge localization is enhanced on the Cr atomic sites, thereby, increasing the magnetic moment and band gap [Table \ref{prevstudies}]. Interestingly, the added Hubbard parameter disrupts the ratio of inter- and intra-layer exchange couplings, thus predicting a X-AFM ground state with all other phases at least $1.5$ eV/Cr higher in energy. 

%Fig 3 bz, bs, dos
\begin{figure*}[ht!]
\centering
\includegraphics[width = 0.99\textwidth]{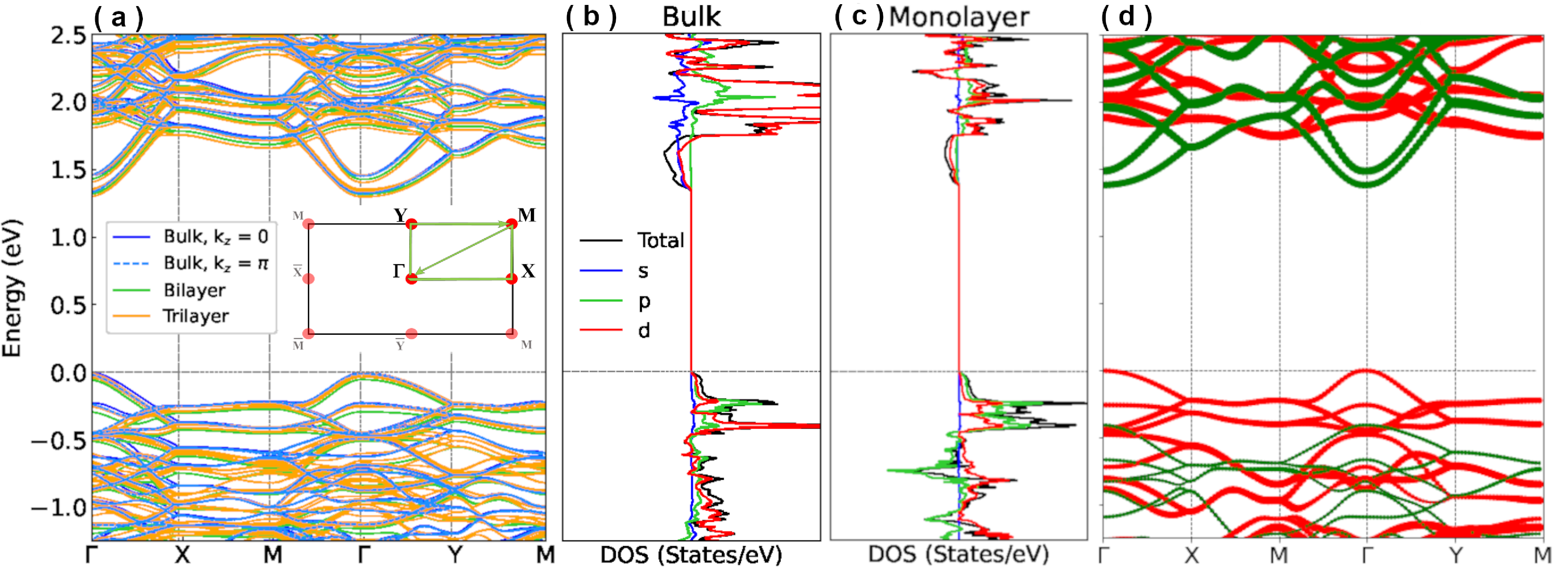}
\caption{(a) The electronic band structure and Brillioun zone route explored (inner panel) for the bulk, bilayer, and trilayer. (b), The total and atomically resolved density of states is related next to the band structure for the bulk. (c) monolayer atomically resolved density of states, with the corresponding (d) monolayer band structure with up (red) and down (green) bands. }
\label{fig:bsdos} 
\end{figure*}

We find the SCAN density functional to remedy the pitfalls of both previous approaches without the introduction of any empirical parameters. Specifically, SCAN recovers the experimental A-AFM ground state ordering and predicts magnetic moments of 3.00 $\mu_B$ tilted off of the $c$-axis by 71.6$^{\circ}$ in excellent accord to the experimential powder and single crystal magnetic neutron diffraction values~\cite{cald:20}. Regarding magnetic moments, the values obtained by neutron scattering involve uncertainties since the chromium form factor is not {\it a priori} known. Note that, when estimating the chromium magnetic moment, we have increased the Wigner–Seitz radius of the integration sphere beyond the default value from 1.323~\AA~(covalent radius) to 1.74~\AA~(3/4 of the Cr–S bond length) in order to fully capture the magnetic density centered on the chromium atomic site and the part originating from hybridization between the chromium and sulfur atoms, see Appendix~\ref{apendix:wsmoment} for more details).

The A-AFM state stabilizes with a band gap of approximately 1.34 eV that opens up around the Fermi energy of the NM system. This gap is in good agreement with optical data~\cite{loui:78}. The physical interpretation of the band gap obtained in the ground-state DFT calculations has been the subject of much debate in the literature over the years. We note that in assessing the physical content of the band gap, one must distinguish between the nature of the effective exchange-correlation potential obtained in the Kohn-Sham (KS) \cite{kohn1965self} and generalized Kohn-Sham (gKS) \cite{seidl1996generalized} formalisms underlying the construction of various functionals. Within the KS scheme one uses the electron density as a variational parameter, thereby yielding KS potentials that are inherently ``multiplicative" and orbital independent. In sharp contrast, the Slater determinate is varied in the gKS formalism, this results in gKS potentials that are orbital dependent and are thus ``nonmultiplicative." In particular, LSDA/GGA band structures involve multiplicative effective potentials, while the current and common SCAN implementations involve nonmultiplicative potentials due to the inclusion of the kinetic energy density as an ingredient, and thus differ in their basic underlying designs. In this connection, Perdew {\it et al.}~\cite{perd:17} have shown that for a given density functional, the gKS band gap is equal to the fundamental band gap in the solid, which is defined as the ground-state energy difference between systems with different numbers of electrons. There is thus a firm basis for comparing computed band gaps within the gKS-based SCAN formalism with the experimentally observed band gaps (excluding excitonic effects). The preceding considerations indicate that as a meta-GGA functional improves the description of the ground state, it will necessarily also lead to improvement in the band gap. To this end, the gKS-based SCAN formalism has demonstrated success in capturing not only the magnetic ground state, but the band gap in a wide array of materials including La$_2$CuO$_4$ \cite{furness2018accurate,lane2018antiferromagnetic,pokharel2022sensitivity}, YBa$_2$Cu$_3$O$_6$ \cite{zhang2020competing}, Sr$_2$IrO$_4$ \cite{lane2020first}, and NiPS$_3$ \cite{lane2020thickness}. In particular, the SCAN ground state has been shown to yield the key optical transitions in La$_2$CuO$_4$ \cite{lane2020landscape} and NiPS$_3$ \cite{lane2022ab}. Therefore, the comprehensive agreement between magnetic and electronic properties in CrPS$_4$ stems from the reduction in self-interaction error in SCAN, as compared to PBE, and is not accidental.

Table~\ref{toten} presents the relative total energy of each magnetic configuration with respect to the A-AFM state for various sample thicknesses. The A-AFM phase is found to be the ground state for all layered-films studied, with a robust FM order stabilizing in the monolayer. By comparing the energy of each magnetic state, the FM phase is the closest competing magnetic state, with only $\approx 2$-$7$ meV/Cr energy separation irrespective of thickness. Curiously, the energy difference between A-AFM and FM arrangements follow a non-monotonic evolution with number of CrPS$_4$ layers. Similarly, the remaining magnetic orders farther away from the ground state display a sensitive dependence on the number of layers. In particular, the X- and Y-AFM phases interchange energetic ordering from the bilayer to the bulk systems. This suggests the screening environment may play a key role in tuning the relative strength between the inter- and intra-chain interactions. Additionally, by comparing C-, G-, X-, and X$^\prime$-AFM configurations, the interlayer exchange coupling is found to be on the order of a few meV/Cr. 

%Fig 4 dielectrics
\begin{figure*}[ht!]
\centering
\includegraphics[width = 0.99\textwidth]{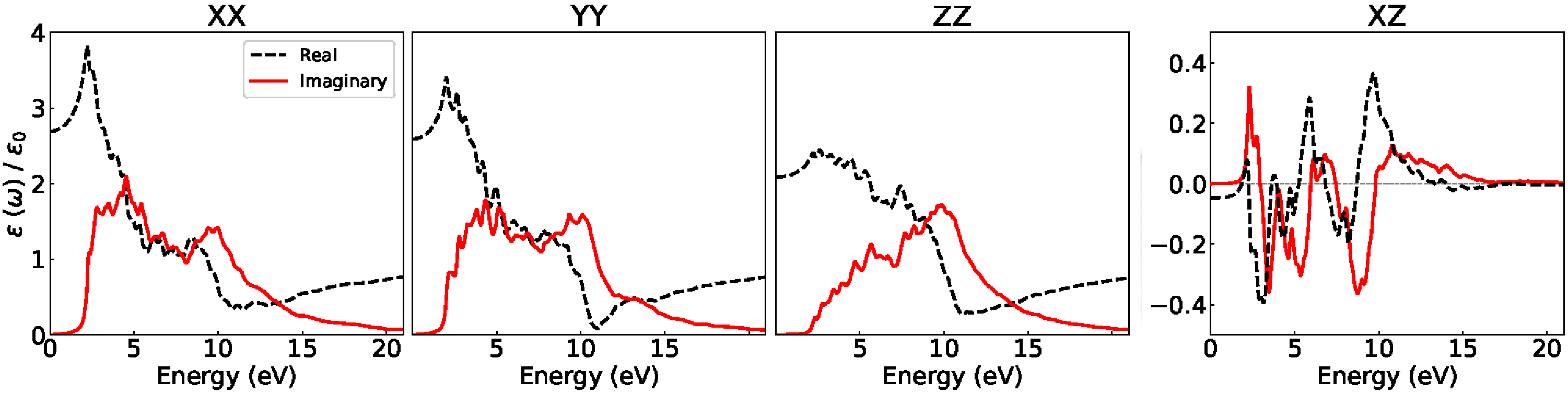}
\caption{Dielectric tensor components of CrPS$_{4}$ FM monolayer showing the real (dashed black line) and imaginary (red) components. The $xy$ and $yz$ components have little to no contribution to the system. }
\label{fig:dielectrics} 
\end{figure*}

%Fig 5 kerr parameters
\begin{figure}[h!]
\centering
\includegraphics[width = 0.65\columnwidth]{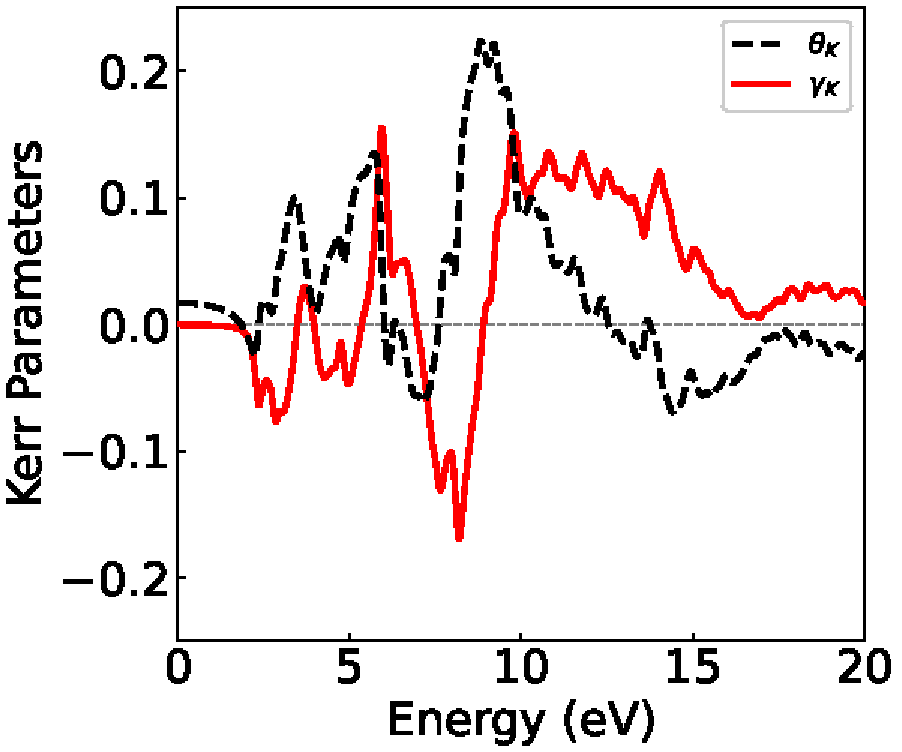}
\caption{The Kerr parameters of Monolayer CrPS$_{4}$ as a function of energy.}
\label{fig:kerr} 
\end{figure}

\section{Electronic Properties}
Figure~\ref{fig:bsdos} shows the (a) electronic band structure and (b) atomically resolved density of states of CrPS$_{4}$ in the A-AFM phase for the bulk, bilayer, and trilayer, with separately the (c) monolayer atomically resolved density of states, and (d) corresponding band structure. For the multi-layer structures, the A-AFM magnetic order stabilizes a direct 1.34 eV band gap at $\Gamma$ in the Brillouin zone of bulk CrPS$_4$. Interestingly, the band dispersions appear to be insensitive to k$_{z}$ dispersion. Specifically, the bands for $k_z=0$ (solid blue lines) and  $k_z=\pi$ (dashed blue lines) are essentially identical, where small deviations are only exhibited along the M-$\Gamma$-Y path in momentum space. The band dispersion along $\Gamma$-Z is flat suggesting the interlayer coupling is weak and each atomically-thin layer is relatively electronically isolated, see Appendix~\ref{apendix:zdisp}. As CrPS$_4$ is thinned, the band structure for the trilayer (orange solid lines) and bilayer (green solid lines) appears to be stationary and display nearly identical eigenvalues throughout the Brillioun zone, except for a slight rigid shift ($\sim 70$ meV) due to the non-monotonic evolution of the gap. In contrast, the monolayer [Fig.~\ref{fig:bsdos}(d)] displays an indirect band gap of 1.745 eV characterized by momentum-shift of $(\pi,\pi)$. The corresponding atomically resolved density of states [Figure~\ref{fig:bsdos}(b) and (c)] reveals the valence bands to be predominantly composed of sulfur character 99\%, whereas the conduction band is dominated by Cr states 27\%. This stacking-of-states follows the Zaanen-Sawatzky-Allen classification of a charge-transfer insulator~\cite{zaan:85}. Therefore, in contrast to the Mott-Hubbard scenario, when a hole is doped into CrPS$_4$, the carrier would sit on the sulfur atomic sites rather than in the Cr sites. The percentage of metal cation (ligend anion) weight in the conduction (valence) band is relatively constant across the bulk, single-, and few-layer systems.

\section{Optical Properties}
To accelerate the design of the next generation microelectronic devices, it is necessary to be able to cross validate both theoretical calculations and experimental measurements. To this end, we first must judge the quality of our theoretically obtained description of CrPS$_4$. To address this issue we calculate the dielectric tensor -- a key ingredient in the interaction between light and matter -- and compare the results to experimental observations.

Figure~\ref{fig:dielectrics} presents the non-zero matrix elements of the dielectric tensor (real and imaginary components) as a function of energy for a single-layer of CrPS$_{4}$. Two main blocks of transitions are seen spanning $\approx 2$ eV - $6$ eV and $\approx 9$ eV - $11$ eV in the imaginary part of the $xx$ and $yy$ tensor components. In contrast, $\varepsilon^{(2)}_{zz}$ displays a very weak leading edge of transitions due to the 2D nature of the material and only retains the broad peak centered at 10 eV. Additionally, the sharp leading edge at $\approx 2$ eV in $\varepsilon^{(2)}$ generates a strong polarization peak in $\varepsilon^{(1)}$ through the Kramer-Kronig relation. On average, the indicated peak structure is in good accord with optical spectroscopy reports~\cite{lee:17}.

Comparing to Fig.~\ref{fig:bsdos}, the sharp transition edge at $\approx 2$ eV is produced by promoting an electron from the valence- to conduction-band edges along X-M (Y-M). The higher energetic transitions originate from bands 5.0 eV below the Fermi level connecting to the flat conduction bands along X-M and Y-M. Furthermore, our theoretically predicted electronic band gap is in very good agreement with the leading edge of the optical conductivity~\cite{loui:78}. To allow for future experimental comparison, the refractive index and optical absorption spectrum are provided in Appendix~\ref{apendix:optical}.

As a consequence of the canted ferromagnetism the off-diagonal $xz$ component of the dielectric tensor exhibits a non-negligible response. Such a response may induce an appreciable Kerr angle in the polarization of the light reflected from a magnetic CrPS$_4$ layer. This makes the optical Kerr effect particularly useful in giving direct insight into the local, microscopic magnetism~\cite{ersk:75} and time-reversal symmetry breaking condensed-matter systems in general~\cite{xia:06}.

To estimate the Kerr response of CrPS$_4$, we compute the complex Kerr parameters according to the equation
\begin{equation}
\ \psi_K = \theta_{K} + i\gamma_{K} = \frac{-\varepsilon_{xz}} {(\varepsilon_{xx} - 1)\sqrt{\varepsilon_{xx}}} ,
\end{equation}
which is the standard expression for the polar geometry in the small angles limit~\cite{sang:12}. Here, the photon propagates along the $y$ direction and describes a linearly polarized wave with the electric field along the $x$ direction.

Figure~\ref{fig:kerr} shows the real and imaginary Kerr parameters $\theta_{K}$ and $\gamma_{K}$, respectively, as a function of frequency. Both Kerr parameters oscillate about zero displaying a quite sensitive dependence on the frequency of light. Specifically, $\theta_{K}$ and $\gamma_{K}$ change sign at $\approx 2$ eV, followed by a number of additional fluctuations about zero until a maximum in both signals is achieved near 9 eV. Above 9 eV the Kerr parameters vary more smoothly with frequency. Additionally, $\theta_{K}$ and $\gamma_{K}$ appear to be completely out of phase for all frequencies studied. Such a line shape in the Kerr signal is quite different from bulk 3d magnets, e.g. Fe, Co, and Ni, and more akin to other 2D magnets, e.g. NiPS$_3$~\cite{lane2020thickness}. 

\section{Concluding Remarks}
Our study has demonstrated that a first-principles parameter free treatment of the ground state magnetic and electronic proprieties of the air stable CrPS$_4$ 2D magnet is possible. Using the SCAN density functional we find a major improvement over standard PBE and PBE+U approaches, at nearly the same computational cost. Our study lays a firm foundation for the predictive exploration and design of new heterostructures and devices composed of CrPS$_4$ mitigating the need for purely experimental efforts. Furthermore, the highly anisotropic cross-coupling between spin, charge, and lattice also provide a path towards multifunctional devices  ideal for monolithically integrating into semiconducting substrates for efficient interfaces between logic – interconnect – memory sectors in emerging 3D logic-memory architectures.

\begin{acknowledgments}
The work at Los Alamos National Laboratory was carried out under the auspices of the US Department of Energy (DOE) National Nuclear Security Administration under Contract No. 89233218CNA000001. It was supported by the LANL LDRD Program, the Quantum Science Center, a U.S. DOE Office of Science National Quantum Information Science Research Center, and in part by the Center for Integrated Nanotechnologies, a DOE BES user facility, in partnership with the LANL Institutional Computing Program for computational resources. Additional computations were performed at the National Energy Research Scientific Computing Center (NERSC), a U.S. Department of Energy Office of Science User Facility located at Lawrence Berkeley National Laboratory, operated under Contract No. DE-AC02-05CH11231 using NERSC award ERCAP0020494.
JTH acknowledges support from the Institute of Materials Science at Los Alamos National Laboratory.
\end{acknowledgments}

\appendix

\section{Chromium Magnetic Moments as a Function of Wigner-Seitz Radius}\label{apendix:wsmoment}
\begin{figure}[h!]
\centering
\includegraphics[width = 0.99\columnwidth]{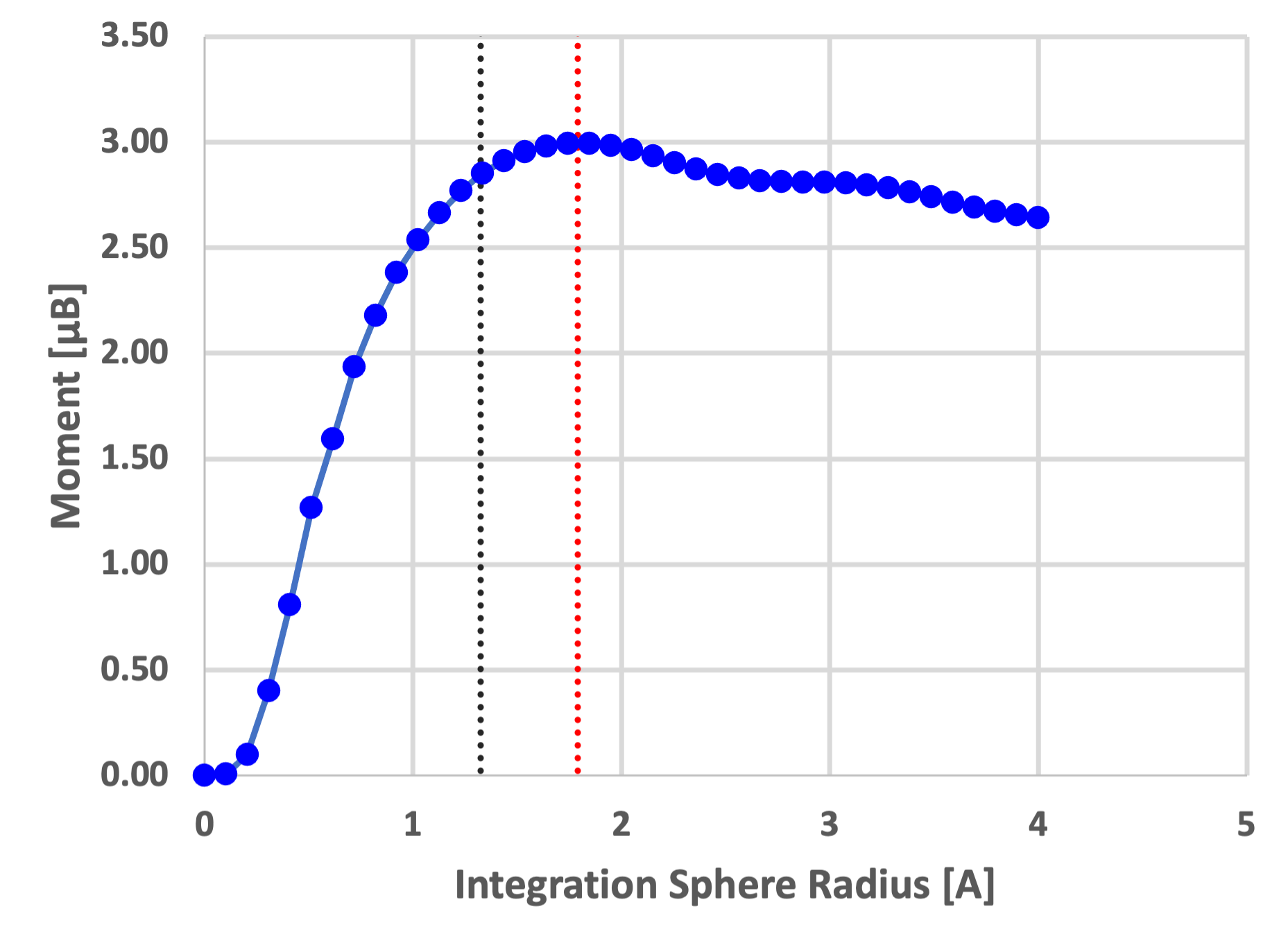}
\caption{(color online) The magnetic moment as a function of integration sphere radius of bulk CrPS$_{4}$ in the A-AFM phase.}
\label{fig:rwigs} 
\end{figure}

Figure~\ref{fig:rwigs} shows the Cr atomic magnetic moment as a function of the Wigner-Seitz radius ($r_{WS}$), or integration sphere. The black dotted line denotes the default value of $r_{WS}$ defined in VASP, which yields a magnetic moment of 2.78~$\mu_B$. Upon expanding $r_{WS}$ we find the magnetic moment to increase, since the integration sphere encapsulates more chromium magnetic density. At $r_{WS}=1.74$~\AA, a maximum in the magnetic moment is found with a value of 3.00~$\mu_B$ (red dotted line). Moreover, as the integration sphere is expanded beyond the covalent radius of chromium, an admixture of sulfur density is included, thereby reducing the net effective moment. Note that the Cr moment for large $r_{WS}$ values will depend on the coordination number and thus on the crystal structure.

\section{$\Gamma$-Z Band Dispersion}\label{apendix:zdisp}
\begin{figure}[th!]
\centering
\includegraphics[width = 0.90\columnwidth]{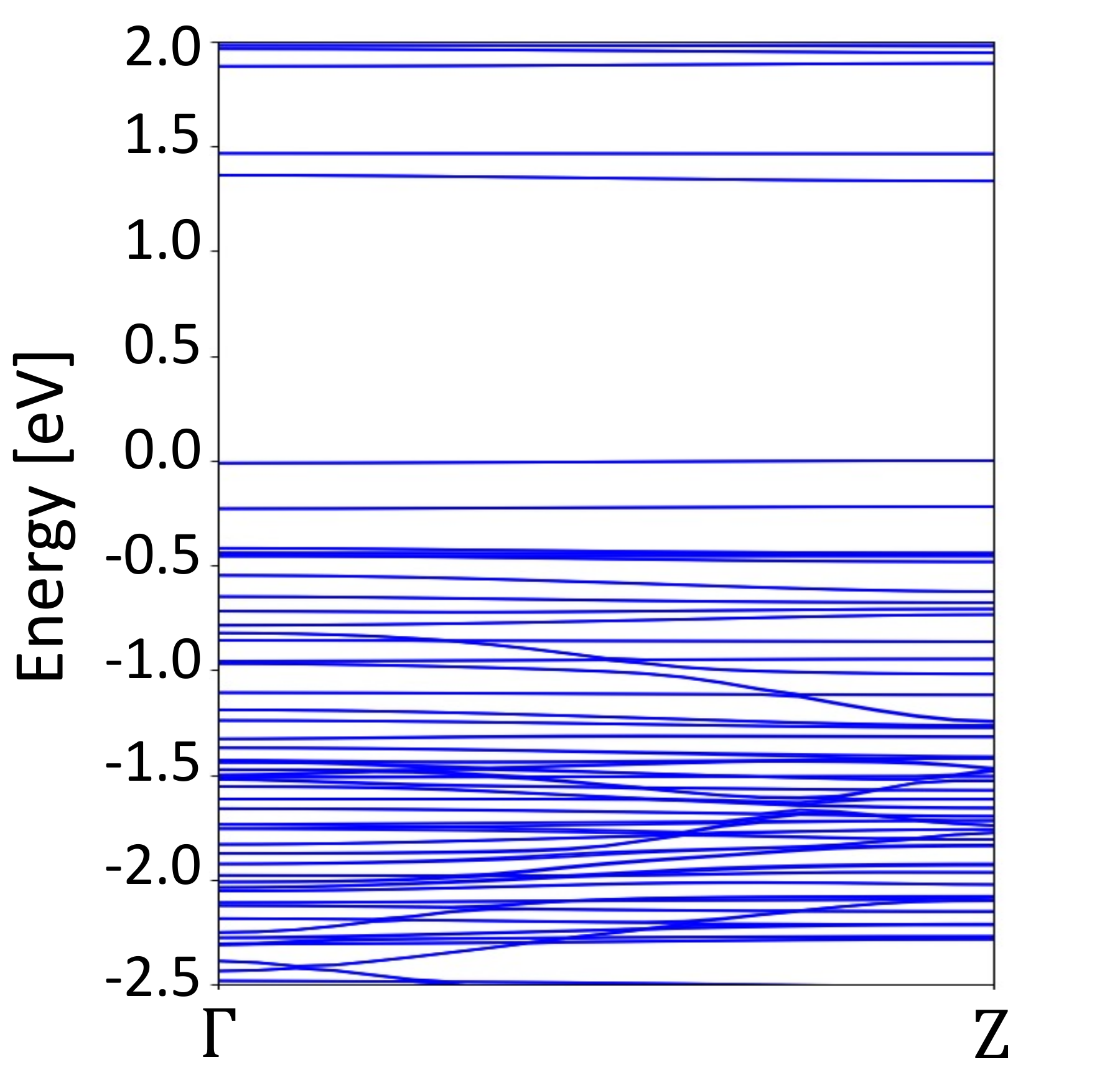}
\caption{(color online) Electronic band structure along $\Gamma$-Z in the Brillouin zone for bulk CrPS$_{4}$ in the A-AFM phase.}
\label{fig:kz} 
\end{figure}

Figure~\ref{fig:kz} shows the electronic band structure along $\Gamma$-Z in the Brillouin zone for bulk CrPS$_{4}$ in the A-AFM phase. The majority of bands are essentially flat, with only a few valence states below -0.75 eV displaying a dispersive nature. This suggests the inter-layer coupling in CrPS$_{4}$ is quite weak, rendering the atomically-thin layers even in the bulk compound electronically isolated. 

\section{Refractive Index and Optical Absorption Spectrum}\label{apendix:optical}
\begin{figure}[th!]
\centering
\includegraphics[width = 0.90\columnwidth]{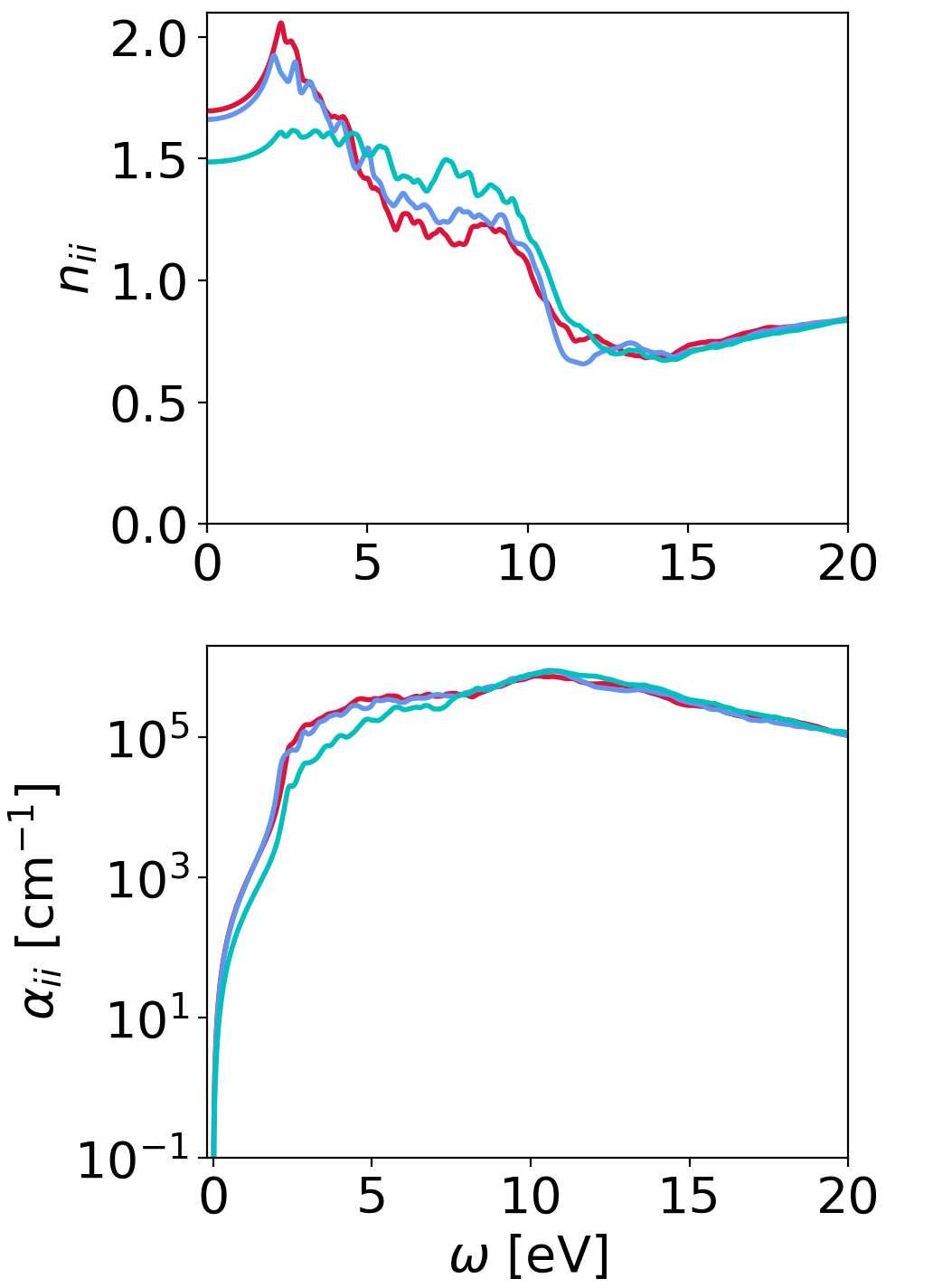}
\caption{(color online) Refractive index and optical absorption for monolayer CrPS$_{4}$ in the FM phase. The red, blue, and teal lines are for the $xx$, $yy$, and $zz$ components, respectively.}
\label{fig:op} 
\end{figure}

Figure~\ref{fig:op} presents the refractive index and optical absorption for monolayer CrPS$_{4}$ in the FM phase. The refractive index $n_{ii}$ is related to the diagonal parts of the dielectric tensor by
\begin{align}
n_{ii} = \sqrt{\frac{|\varepsilon_{ii}|+\Re{\varepsilon}_{ii}}{2}},
\end{align}
whereas the absorption coefficient 
\begin{align}
\alpha_{ii}=\frac{2\omega k_{ii}}{c},
\end{align}
is proportional to the dielectric tensor through extinction coefficient,
\begin{align}
k_{ii} = \sqrt{\frac{|\varepsilon_{ii}|-\Re{\varepsilon}_{ii}}{2}}.
\end{align}

\bibliography{manuscript}

\begin{thebibliography}{48}%
\makeatletter
\providecommand \@ifxundefined [1]{%
 \@ifx{#1\undefined}
}%
\providecommand \@ifnum [1]{%
 \ifnum #1\expandafter \@firstoftwo
 \else \expandafter \@secondoftwo
 \fi
}%
\providecommand \@ifx [1]{%
 \ifx #1\expandafter \@firstoftwo
 \else \expandafter \@secondoftwo
 \fi
}%
\providecommand \natexlab [1]{#1}%
\providecommand \enquote  [1]{``#1''}%
\providecommand \bibnamefont  [1]{#1}%
\providecommand \bibfnamefont [1]{#1}%
\providecommand \citenamefont [1]{#1}%
\providecommand \href@noop [0]{\@secondoftwo}%
\providecommand \href [0]{\begingroup \@sanitize@url \@href}%
\providecommand \@href[1]{\@@startlink{#1}\@@href}%
\providecommand \@@href[1]{\endgroup#1\@@endlink}%
\providecommand \@sanitize@url [0]{\catcode `\\12\catcode `\$12\catcode
  `\&12\catcode `\#12\catcode `\^12\catcode `\_12\catcode `\%12\relax}%
\providecommand \@@startlink[1]{}%
\providecommand \@@endlink[0]{}%
\providecommand \url  [0]{\begingroup\@sanitize@url \@url }%
\providecommand \@url [1]{\endgroup\@href {#1}{\urlprefix }}%
\providecommand \urlprefix  [0]{URL }%
\providecommand \Eprint [0]{\href }%
\providecommand \doibase [0]{https://doi.org/}%
\providecommand \selectlanguage [0]{\@gobble}%
\providecommand \bibinfo  [0]{\@secondoftwo}%
\providecommand \bibfield  [0]{\@secondoftwo}%
\providecommand \translation [1]{[#1]}%
\providecommand \BibitemOpen [0]{}%
\providecommand \bibitemStop [0]{}%
\providecommand \bibitemNoStop [0]{.\EOS\space}%
\providecommand \EOS [0]{\spacefactor3000\relax}%
\providecommand \BibitemShut  [1]{\csname bibitem#1\endcsname}%
\let\auto@bib@innerbib\@empty
%</preamble>
\bibitem [{\citenamefont {Gibertini}\ \emph {et~al.}(2019)\citenamefont
  {Gibertini}, \citenamefont {Koperski}, \citenamefont {Morpurgo},\ and\
  \citenamefont {Novoselov}}]{gibe:19}%
  \BibitemOpen
  \bibfield  {author} {\bibinfo {author} {\bibfnamefont {M.}~\bibnamefont
  {Gibertini}}, \bibinfo {author} {\bibfnamefont {M.}~\bibnamefont {Koperski}},
  \bibinfo {author} {\bibfnamefont {A.}~\bibnamefont {Morpurgo}},\ and\
  \bibinfo {author} {\bibfnamefont {K.}~\bibnamefont {Novoselov}},\ }\bibfield
  {title} {\bibinfo {title} {{Magnetic 2D materials and heterostructures}},\
  }\href@noop {} {\bibfield  {journal} {\bibinfo  {journal} {Nat.
  Nanotechnol.}\ }\textbf {\bibinfo {volume} {14}},\ \bibinfo {pages} {408}
  (\bibinfo {year} {2019})}\BibitemShut {NoStop}%
\bibitem [{\citenamefont {Li}\ \emph {et~al.}(2021)\citenamefont {Li},
  \citenamefont {Zaki}, \citenamefont {Garlea}, \citenamefont {Savici},
  \citenamefont {Fobes}, \citenamefont {Xu}, \citenamefont {Camino},
  \citenamefont {Petrovic}, \citenamefont {Gu}, \citenamefont {Johnson} \emph
  {et~al.}}]{li2021electronic}%
  \BibitemOpen
  \bibfield  {author} {\bibinfo {author} {\bibfnamefont {Y.}~\bibnamefont
  {Li}}, \bibinfo {author} {\bibfnamefont {N.}~\bibnamefont {Zaki}}, \bibinfo
  {author} {\bibfnamefont {V.~O.}\ \bibnamefont {Garlea}}, \bibinfo {author}
  {\bibfnamefont {A.~T.}\ \bibnamefont {Savici}}, \bibinfo {author}
  {\bibfnamefont {D.}~\bibnamefont {Fobes}}, \bibinfo {author} {\bibfnamefont
  {Z.}~\bibnamefont {Xu}}, \bibinfo {author} {\bibfnamefont {F.}~\bibnamefont
  {Camino}}, \bibinfo {author} {\bibfnamefont {C.}~\bibnamefont {Petrovic}},
  \bibinfo {author} {\bibfnamefont {G.}~\bibnamefont {Gu}}, \bibinfo {author}
  {\bibfnamefont {P.~D.}\ \bibnamefont {Johnson}}, \emph {et~al.},\ }\bibfield
  {title} {\bibinfo {title} {Electronic properties of the bulk and surface
  states of fe$_{1+ y}$te$_{1- x}$se$_{x}$},\ }\href@noop {} {\bibfield
  {journal} {\bibinfo  {journal} {Nature Materials}\ }\textbf {\bibinfo
  {volume} {20}},\ \bibinfo {pages} {1221} (\bibinfo {year}
  {2021})}\BibitemShut {NoStop}%
\bibitem [{\citenamefont {Kim}\ \emph {et~al.}(2018)\citenamefont {Kim},
  \citenamefont {Seo}, \citenamefont {Lee}, \citenamefont {Ko}, \citenamefont
  {Kim}, \citenamefont {Jang}, \citenamefont {Ok}, \citenamefont {Lee},
  \citenamefont {Jo}, \citenamefont {Kang} \emph {et~al.}}]{kim2018large}%
  \BibitemOpen
  \bibfield  {author} {\bibinfo {author} {\bibfnamefont {K.}~\bibnamefont
  {Kim}}, \bibinfo {author} {\bibfnamefont {J.}~\bibnamefont {Seo}}, \bibinfo
  {author} {\bibfnamefont {E.}~\bibnamefont {Lee}}, \bibinfo {author}
  {\bibfnamefont {K.-T.}\ \bibnamefont {Ko}}, \bibinfo {author} {\bibfnamefont
  {B.}~\bibnamefont {Kim}}, \bibinfo {author} {\bibfnamefont {B.~G.}\
  \bibnamefont {Jang}}, \bibinfo {author} {\bibfnamefont {J.~M.}\ \bibnamefont
  {Ok}}, \bibinfo {author} {\bibfnamefont {J.}~\bibnamefont {Lee}}, \bibinfo
  {author} {\bibfnamefont {Y.~J.}\ \bibnamefont {Jo}}, \bibinfo {author}
  {\bibfnamefont {W.}~\bibnamefont {Kang}}, \emph {et~al.},\ }\bibfield
  {title} {\bibinfo {title} {Large anomalous hall current induced by
  topological nodal lines in a ferromagnetic van der waals semimetal},\
  }\href@noop {} {\bibfield  {journal} {\bibinfo  {journal} {Nature Materials}\
  }\textbf {\bibinfo {volume} {17}},\ \bibinfo {pages} {794} (\bibinfo {year}
  {2018})}\BibitemShut {NoStop}%
\bibitem [{\citenamefont {Burch}\ \emph {et~al.}(2018)\citenamefont {Burch},
  \citenamefont {Mandrus},\ and\ \citenamefont {Park}}]{burch2018magnetism}%
  \BibitemOpen
  \bibfield  {author} {\bibinfo {author} {\bibfnamefont {K.~S.}\ \bibnamefont
  {Burch}}, \bibinfo {author} {\bibfnamefont {D.}~\bibnamefont {Mandrus}},\
  and\ \bibinfo {author} {\bibfnamefont {J.-G.}\ \bibnamefont {Park}},\
  }\bibfield  {title} {\bibinfo {title} {Magnetism in two-dimensional van der
  waals materials},\ }\href@noop {} {\bibfield  {journal} {\bibinfo  {journal}
  {Nature}\ }\textbf {\bibinfo {volume} {563}},\ \bibinfo {pages} {47}
  (\bibinfo {year} {2018})}\BibitemShut {NoStop}%
\bibitem [{\citenamefont {Shcherbakov}\ \emph {et~al.}(2018)\citenamefont
  {Shcherbakov}, \citenamefont {Stepanov}, \citenamefont {Weber}, \citenamefont
  {Wang}, \citenamefont {Hu}, \citenamefont {Zhu}, \citenamefont {Watanabe},
  \citenamefont {Taniguchi}, \citenamefont {Mao}, \citenamefont {Windl} \emph
  {et~al.}}]{shcherbakov2018raman}%
  \BibitemOpen
  \bibfield  {author} {\bibinfo {author} {\bibfnamefont {D.}~\bibnamefont
  {Shcherbakov}}, \bibinfo {author} {\bibfnamefont {P.}~\bibnamefont
  {Stepanov}}, \bibinfo {author} {\bibfnamefont {D.}~\bibnamefont {Weber}},
  \bibinfo {author} {\bibfnamefont {Y.}~\bibnamefont {Wang}}, \bibinfo {author}
  {\bibfnamefont {J.}~\bibnamefont {Hu}}, \bibinfo {author} {\bibfnamefont
  {Y.}~\bibnamefont {Zhu}}, \bibinfo {author} {\bibfnamefont {K.}~\bibnamefont
  {Watanabe}}, \bibinfo {author} {\bibfnamefont {T.}~\bibnamefont {Taniguchi}},
  \bibinfo {author} {\bibfnamefont {Z.}~\bibnamefont {Mao}}, \bibinfo {author}
  {\bibfnamefont {W.}~\bibnamefont {Windl}}, \emph {et~al.},\ }\bibfield
  {title} {\bibinfo {title} {Raman spectroscopy, photocatalytic degradation,
  and stabilization of atomically thin chromium tri-iodide},\ }\href@noop {}
  {\bibfield  {journal} {\bibinfo  {journal} {Nano Letters}\ }\textbf {\bibinfo
  {volume} {18}},\ \bibinfo {pages} {4214} (\bibinfo {year}
  {2018})}\BibitemShut {NoStop}%
\bibitem [{\citenamefont {Calder}\ \emph {et~al.}(2020)\citenamefont {Calder},
  \citenamefont {Haglund}, \citenamefont {Liu}, \citenamefont {Pajerowski},
  \citenamefont {Cau}, \citenamefont {Williams}, \citenamefont {Garlea},\ and\
  \citenamefont {Mandrus}}]{cald:20}%
  \BibitemOpen
  \bibfield  {author} {\bibinfo {author} {\bibfnamefont {S.}~\bibnamefont
  {Calder}}, \bibinfo {author} {\bibfnamefont {A.}~\bibnamefont {Haglund}},
  \bibinfo {author} {\bibfnamefont {Y.}~\bibnamefont {Liu}}, \bibinfo {author}
  {\bibfnamefont {D.}~\bibnamefont {Pajerowski}}, \bibinfo {author}
  {\bibfnamefont {H.}~\bibnamefont {Cau}}, \bibinfo {author} {\bibfnamefont
  {T.}~\bibnamefont {Williams}}, \bibinfo {author} {\bibfnamefont
  {V.}~\bibnamefont {Garlea}},\ and\ \bibinfo {author} {\bibfnamefont
  {D.}~\bibnamefont {Mandrus}},\ }\bibfield  {title} {\bibinfo {title}
  {{Magnetic structure and exchange interactions in the layered semiconductor
  CrPS$_{4}$}},\ }\href@noop {} {\bibfield  {journal} {\bibinfo  {journal}
  {Physical Review B}\ }\textbf {\bibinfo {volume} {102}},\ \bibinfo {pages}
  {024408} (\bibinfo {year} {2020})}\BibitemShut {NoStop}%
\bibitem [{\citenamefont {Son}\ \emph {et~al.}(2021)\citenamefont {Son},
  \citenamefont {Son}, \citenamefont {Park}, \citenamefont {Kim}, \citenamefont
  {Tao}, \citenamefont {Oh}, \citenamefont {Lee}, \citenamefont {Lee},
  \citenamefont {Kim}, \citenamefont {Zhang}, \citenamefont {Cho},
  \citenamefont {Kamiyama}, \citenamefont {Lee}, \citenamefont {Mak},
  \citenamefont {Shan}, \citenamefont {Kim}, \citenamefont {Park},\ and\
  \citenamefont {Lee}}]{sonj:21}%
  \BibitemOpen
  \bibfield  {author} {\bibinfo {author} {\bibfnamefont {J.}~\bibnamefont
  {Son}}, \bibinfo {author} {\bibfnamefont {S.}~\bibnamefont {Son}}, \bibinfo
  {author} {\bibfnamefont {P.}~\bibnamefont {Park}}, \bibinfo {author}
  {\bibfnamefont {M.}~\bibnamefont {Kim}}, \bibinfo {author} {\bibfnamefont
  {Z.}~\bibnamefont {Tao}}, \bibinfo {author} {\bibfnamefont {J.}~\bibnamefont
  {Oh}}, \bibinfo {author} {\bibfnamefont {T.}~\bibnamefont {Lee}}, \bibinfo
  {author} {\bibfnamefont {S.}~\bibnamefont {Lee}}, \bibinfo {author}
  {\bibfnamefont {J.}~\bibnamefont {Kim}}, \bibinfo {author} {\bibfnamefont
  {K.}~\bibnamefont {Zhang}}, \bibinfo {author} {\bibfnamefont
  {K.}~\bibnamefont {Cho}}, \bibinfo {author} {\bibfnamefont {T.}~\bibnamefont
  {Kamiyama}}, \bibinfo {author} {\bibfnamefont {J.}~\bibnamefont {Lee}},
  \bibinfo {author} {\bibfnamefont {K.}~\bibnamefont {Mak}}, \bibinfo {author}
  {\bibfnamefont {J.}~\bibnamefont {Shan}}, \bibinfo {author} {\bibfnamefont
  {M.}~\bibnamefont {Kim}}, \bibinfo {author} {\bibfnamefont {J.}~\bibnamefont
  {Park}},\ and\ \bibinfo {author} {\bibfnamefont {J.}~\bibnamefont {Lee}},\
  }\bibfield  {title} {\bibinfo {title} {{Air-Stable and Layer-Dependent
  Ferromagnetism in Atomically Thin van der Waals CrPS$_{4}$}},\ }\href@noop {}
  {\bibfield  {journal} {\bibinfo  {journal} {ACS Nano}\ }\textbf {\bibinfo
  {volume} {15}},\ \bibinfo {pages} {16904} (\bibinfo {year}
  {2021})}\BibitemShut {NoStop}%
\bibitem [{\citenamefont {Meng}\ \emph {et~al.}(2021)\citenamefont {Meng},
  \citenamefont {Zhou}, \citenamefont {Xu}, \citenamefont {Yang}, \citenamefont
  {Si}, \citenamefont {Liu}, \citenamefont {Wang}, \citenamefont {Jiang},
  \citenamefont {Li}, \citenamefont {Qin} \emph {et~al.}}]{meng2021anomalous}%
  \BibitemOpen
  \bibfield  {author} {\bibinfo {author} {\bibfnamefont {L.}~\bibnamefont
  {Meng}}, \bibinfo {author} {\bibfnamefont {Z.}~\bibnamefont {Zhou}}, \bibinfo
  {author} {\bibfnamefont {M.}~\bibnamefont {Xu}}, \bibinfo {author}
  {\bibfnamefont {S.}~\bibnamefont {Yang}}, \bibinfo {author} {\bibfnamefont
  {K.}~\bibnamefont {Si}}, \bibinfo {author} {\bibfnamefont {L.}~\bibnamefont
  {Liu}}, \bibinfo {author} {\bibfnamefont {X.}~\bibnamefont {Wang}}, \bibinfo
  {author} {\bibfnamefont {H.}~\bibnamefont {Jiang}}, \bibinfo {author}
  {\bibfnamefont {B.}~\bibnamefont {Li}}, \bibinfo {author} {\bibfnamefont
  {P.}~\bibnamefont {Qin}}, \emph {et~al.},\ }\bibfield  {title} {\bibinfo
  {title} {Anomalous thickness dependence of curie temperature in air-stable
  two-dimensional ferromagnetic 1t-crte2 grown by chemical vapor deposition},\
  }\href@noop {} {\bibfield  {journal} {\bibinfo  {journal} {Nature
  Communications}\ }\textbf {\bibinfo {volume} {12}},\ \bibinfo {pages} {809}
  (\bibinfo {year} {2021})}\BibitemShut {NoStop}%
\bibitem [{\citenamefont {Purbawati}\ \emph {et~al.}(2023)\citenamefont
  {Purbawati}, \citenamefont {Sarkar}, \citenamefont {Pairis}, \citenamefont
  {Kostka}, \citenamefont {Hadj-Azzem}, \citenamefont {Dufeu}, \citenamefont
  {Singh}, \citenamefont {Bourgault}, \citenamefont {Nu{\~n}ez-Regueiro},
  \citenamefont {Vogel} \emph {et~al.}}]{purbawati2023stability}%
  \BibitemOpen
  \bibfield  {author} {\bibinfo {author} {\bibfnamefont {A.}~\bibnamefont
  {Purbawati}}, \bibinfo {author} {\bibfnamefont {S.}~\bibnamefont {Sarkar}},
  \bibinfo {author} {\bibfnamefont {S.}~\bibnamefont {Pairis}}, \bibinfo
  {author} {\bibfnamefont {M.}~\bibnamefont {Kostka}}, \bibinfo {author}
  {\bibfnamefont {A.}~\bibnamefont {Hadj-Azzem}}, \bibinfo {author}
  {\bibfnamefont {D.}~\bibnamefont {Dufeu}}, \bibinfo {author} {\bibfnamefont
  {P.}~\bibnamefont {Singh}}, \bibinfo {author} {\bibfnamefont
  {D.}~\bibnamefont {Bourgault}}, \bibinfo {author} {\bibfnamefont
  {M.}~\bibnamefont {Nu{\~n}ez-Regueiro}}, \bibinfo {author} {\bibfnamefont
  {J.}~\bibnamefont {Vogel}}, \emph {et~al.},\ }\bibfield  {title} {\bibinfo
  {title} {Stability of the in-plane room temperature van der waals ferromagnet
  chromium ditelluride and its conversion to chromium-interleaved crte2
  compounds},\ }\href@noop {} {\bibfield  {journal} {\bibinfo  {journal} {ACS
  Applied Electronic Materials}\ } (\bibinfo {year} {2023})}\BibitemShut
  {NoStop}%
\bibitem [{\citenamefont {Huey}\ \emph {et~al.}(2022)\citenamefont {Huey},
  \citenamefont {Ochs}, \citenamefont {Williams}, \citenamefont {Zhang},
  \citenamefont {Kraguljac}, \citenamefont {Deng}, \citenamefont {Moore},
  \citenamefont {Windl}, \citenamefont {Lau},\ and\ \citenamefont
  {Goldberger}}]{huey2022cr}%
  \BibitemOpen
  \bibfield  {author} {\bibinfo {author} {\bibfnamefont {W.~L.}\ \bibnamefont
  {Huey}}, \bibinfo {author} {\bibfnamefont {A.~M.}\ \bibnamefont {Ochs}},
  \bibinfo {author} {\bibfnamefont {A.~J.}\ \bibnamefont {Williams}}, \bibinfo
  {author} {\bibfnamefont {Y.}~\bibnamefont {Zhang}}, \bibinfo {author}
  {\bibfnamefont {S.}~\bibnamefont {Kraguljac}}, \bibinfo {author}
  {\bibfnamefont {Z.}~\bibnamefont {Deng}}, \bibinfo {author} {\bibfnamefont
  {C.~E.}\ \bibnamefont {Moore}}, \bibinfo {author} {\bibfnamefont
  {W.}~\bibnamefont {Windl}}, \bibinfo {author} {\bibfnamefont {C.~N.}\
  \bibnamefont {Lau}},\ and\ \bibinfo {author} {\bibfnamefont {J.~E.}\
  \bibnamefont {Goldberger}},\ }\bibfield  {title} {\bibinfo {title} {Cr x
  pt1--x te2 ($x\le$0.45): A family of air-stable and exfoliatable van der
  waals ferromagnets},\ }\href@noop {} {\bibfield  {journal} {\bibinfo
  {journal} {ACS nano}\ }\textbf {\bibinfo {volume} {16}},\ \bibinfo {pages}
  {3852} (\bibinfo {year} {2022})}\BibitemShut {NoStop}%
\bibitem [{\citenamefont {Pei}\ \emph {et~al.}(2016)\citenamefont {Pei},
  \citenamefont {Luo}, \citenamefont {Lin}, \citenamefont {Song}, \citenamefont
  {Hu}, \citenamefont {Zou}, \citenamefont {Yu}, \citenamefont {Yong},
  \citenamefont {Song}, \citenamefont {Lu},\ and\ \citenamefont
  {Sun}}]{peiq:16}%
  \BibitemOpen
  \bibfield  {author} {\bibinfo {author} {\bibfnamefont {Q.}~\bibnamefont
  {Pei}}, \bibinfo {author} {\bibfnamefont {X.}~\bibnamefont {Luo}}, \bibinfo
  {author} {\bibfnamefont {G.}~\bibnamefont {Lin}}, \bibinfo {author}
  {\bibfnamefont {J.}~\bibnamefont {Song}}, \bibinfo {author} {\bibfnamefont
  {L.}~\bibnamefont {Hu}}, \bibinfo {author} {\bibfnamefont {Y.}~\bibnamefont
  {Zou}}, \bibinfo {author} {\bibfnamefont {L.}~\bibnamefont {Yu}}, \bibinfo
  {author} {\bibfnamefont {W.}~\bibnamefont {Yong}}, \bibinfo {author}
  {\bibfnamefont {W.}~\bibnamefont {Song}}, \bibinfo {author} {\bibfnamefont
  {W.}~\bibnamefont {Lu}},\ and\ \bibinfo {author} {\bibfnamefont
  {Y.}~\bibnamefont {Sun}},\ }\bibfield  {title} {\bibinfo {title} {{Spin
  dynamics, electronic, and thermal transport properties of two-dimensional
  CrPS$_{4}$ single crystal}},\ }\href@noop {} {\bibfield  {journal} {\bibinfo
  {journal} {Journal of Applied Physics}\ }\textbf {\bibinfo {volume} {119}},\
  \bibinfo {pages} {043902} (\bibinfo {year} {2016})}\BibitemShut {NoStop}%
\bibitem [{\citenamefont {Lee}\ \emph {et~al.}(2017)\citenamefont {Lee},
  \citenamefont {Ko}, \citenamefont {Kim}, \citenamefont {Bark}, \citenamefont
  {Kang}, \citenamefont {Jung}, \citenamefont {Park}, \citenamefont {Lee},
  \citenamefont {Ryu},\ and\ \citenamefont {Lee}}]{lee:17}%
  \BibitemOpen
  \bibfield  {author} {\bibinfo {author} {\bibfnamefont {J.}~\bibnamefont
  {Lee}}, \bibinfo {author} {\bibfnamefont {T.}~\bibnamefont {Ko}}, \bibinfo
  {author} {\bibfnamefont {J.}~\bibnamefont {Kim}}, \bibinfo {author}
  {\bibfnamefont {H.}~\bibnamefont {Bark}}, \bibinfo {author} {\bibfnamefont
  {B.}~\bibnamefont {Kang}}, \bibinfo {author} {\bibfnamefont {S.}~\bibnamefont
  {Jung}}, \bibinfo {author} {\bibfnamefont {T.}~\bibnamefont {Park}}, \bibinfo
  {author} {\bibfnamefont {Z.}~\bibnamefont {Lee}}, \bibinfo {author}
  {\bibfnamefont {S.}~\bibnamefont {Ryu}},\ and\ \bibinfo {author}
  {\bibfnamefont {C.}~\bibnamefont {Lee}},\ }\bibfield  {title} {\bibinfo
  {title} {{Structural and Optical Properties of Single- and Few-layer Magnetic
  Semiconductor CrPS$_{4}$}},\ }\href@noop {} {\bibfield  {journal} {\bibinfo
  {journal} {ACS Nano}\ }\textbf {\bibinfo {volume} {11}},\ \bibinfo {pages}
  {10935} (\bibinfo {year} {2017})}\BibitemShut {NoStop}%
\bibitem [{\citenamefont {Budniak}\ \emph {et~al.}(2020)\citenamefont
  {Budniak}, \citenamefont {Killileaw}, \citenamefont {Zelewski}, \citenamefont
  {Syntnyk}, \citenamefont {Kauffmann}, \citenamefont {Amouyal}, \citenamefont
  {Kudrawiec}, \citenamefont {Heiss},\ and\ \citenamefont
  {Lifshitz}}]{budn:20}%
  \BibitemOpen
  \bibfield  {author} {\bibinfo {author} {\bibfnamefont {A.}~\bibnamefont
  {Budniak}}, \bibinfo {author} {\bibfnamefont {N.}~\bibnamefont {Killileaw}},
  \bibinfo {author} {\bibfnamefont {S.}~\bibnamefont {Zelewski}}, \bibinfo
  {author} {\bibfnamefont {M.}~\bibnamefont {Syntnyk}}, \bibinfo {author}
  {\bibfnamefont {Y.}~\bibnamefont {Kauffmann}}, \bibinfo {author}
  {\bibfnamefont {Y.}~\bibnamefont {Amouyal}}, \bibinfo {author} {\bibfnamefont
  {R.}~\bibnamefont {Kudrawiec}}, \bibinfo {author} {\bibfnamefont
  {W.}~\bibnamefont {Heiss}},\ and\ \bibinfo {author} {\bibfnamefont
  {E.}~\bibnamefont {Lifshitz}},\ }\bibfield  {title} {\bibinfo {title}
  {{Exfoliated CrPS$_{4}$ with Promising Photoconductivity}},\ }\href@noop {}
  {\bibfield  {journal} {\bibinfo  {journal} {Nano Micro Small}\ }\textbf
  {\bibinfo {volume} {16}},\ \bibinfo {pages} {1905924} (\bibinfo {year}
  {2020})}\BibitemShut {NoStop}%
\bibitem [{\citenamefont {Gu}\ \emph {et~al.}(2019)\citenamefont {Gu},
  \citenamefont {Tan}, \citenamefont {Wan}, \citenamefont {Li}, \citenamefont
  {Peng}, \citenamefont {Lai}, \citenamefont {Ma}, \citenamefont {Yao},
  \citenamefont {Yang}, \citenamefont {Yuan} \emph {et~al.}}]{gupi:19}%
  \BibitemOpen
  \bibfield  {author} {\bibinfo {author} {\bibfnamefont {P.}~\bibnamefont
  {Gu}}, \bibinfo {author} {\bibfnamefont {Q.}~\bibnamefont {Tan}}, \bibinfo
  {author} {\bibfnamefont {Y.}~\bibnamefont {Wan}}, \bibinfo {author}
  {\bibfnamefont {Z.}~\bibnamefont {Li}}, \bibinfo {author} {\bibfnamefont
  {Y.}~\bibnamefont {Peng}}, \bibinfo {author} {\bibfnamefont {J.}~\bibnamefont
  {Lai}}, \bibinfo {author} {\bibfnamefont {J.}~\bibnamefont {Ma}}, \bibinfo
  {author} {\bibfnamefont {X.}~\bibnamefont {Yao}}, \bibinfo {author}
  {\bibfnamefont {S.}~\bibnamefont {Yang}}, \bibinfo {author} {\bibfnamefont
  {K.}~\bibnamefont {Yuan}}, \emph {et~al.},\ }\bibfield  {title} {\bibinfo
  {title} {Photoluminescent quantum interference in a van der waals magnet
  preserved by symmetry breaking},\ }\href@noop {} {\bibfield  {journal}
  {\bibinfo  {journal} {ACS Nano}\ }\textbf {\bibinfo {volume} {14}},\ \bibinfo
  {pages} {1003} (\bibinfo {year} {2019})}\BibitemShut {NoStop}%
\bibitem [{\citenamefont {Kim}\ \emph {et~al.}(2019)\citenamefont {Kim},
  \citenamefont {Lee}, \citenamefont {Jin}, \citenamefont {Jo}, \citenamefont
  {Lee},\ and\ \citenamefont {Ryu}}]{kim:19}%
  \BibitemOpen
  \bibfield  {author} {\bibinfo {author} {\bibfnamefont {S.}~\bibnamefont
  {Kim}}, \bibinfo {author} {\bibfnamefont {J.}~\bibnamefont {Lee}}, \bibinfo
  {author} {\bibfnamefont {G.}~\bibnamefont {Jin}}, \bibinfo {author}
  {\bibfnamefont {M.}~\bibnamefont {Jo}}, \bibinfo {author} {\bibfnamefont
  {C.}~\bibnamefont {Lee}},\ and\ \bibinfo {author} {\bibfnamefont
  {S.}~\bibnamefont {Ryu}},\ }\bibfield  {title} {\bibinfo {title} {{Crossover
  between Photochemical and Photothermal Oxidations of Atomically Thin Magnetic
  Semiconductor CrPS$_{4}$}},\ }\href@noop {} {\bibfield  {journal} {\bibinfo
  {journal} {Nano Letters}\ }\textbf {\bibinfo {volume} {19}},\ \bibinfo
  {pages} {4043} (\bibinfo {year} {2019})}\BibitemShut {NoStop}%
\bibitem [{\citenamefont {Kim}\ \emph {et~al.}(2021)\citenamefont {Kim},
  \citenamefont {Lee}, \citenamefont {Lee},\ and\ \citenamefont
  {Ryu}}]{kim:21}%
  \BibitemOpen
  \bibfield  {author} {\bibinfo {author} {\bibfnamefont {S.}~\bibnamefont
  {Kim}}, \bibinfo {author} {\bibfnamefont {J.}~\bibnamefont {Lee}}, \bibinfo
  {author} {\bibfnamefont {C.}~\bibnamefont {Lee}},\ and\ \bibinfo {author}
  {\bibfnamefont {S.}~\bibnamefont {Ryu}},\ }\bibfield  {title} {\bibinfo
  {title} {Polarized raman spectra and complex raman tensors of
  antiferromagnetic semiconductor crps$_4$},\ }\href@noop {} {\bibfield
  {journal} {\bibinfo  {journal} {The Journal of Physical Chemistry C}\
  }\textbf {\bibinfo {volume} {125}},\ \bibinfo {pages} {2691} (\bibinfo {year}
  {2021})}\BibitemShut {NoStop}%
\bibitem [{\citenamefont {Zhang}\ \emph {et~al.}(2021)\citenamefont {Zhang},
  \citenamefont {Li}, \citenamefont {Hu}, \citenamefont {Xu}, \citenamefont
  {Chen}, \citenamefont {Li}, \citenamefont {Yin}, \citenamefont {Chen},
  \citenamefont {Tan}, \citenamefont {Kan} \emph {et~al.}}]{zhan:21}%
  \BibitemOpen
  \bibfield  {author} {\bibinfo {author} {\bibfnamefont {H.}~\bibnamefont
  {Zhang}}, \bibinfo {author} {\bibfnamefont {Y.}~\bibnamefont {Li}}, \bibinfo
  {author} {\bibfnamefont {X.}~\bibnamefont {Hu}}, \bibinfo {author}
  {\bibfnamefont {J.}~\bibnamefont {Xu}}, \bibinfo {author} {\bibfnamefont
  {L.}~\bibnamefont {Chen}}, \bibinfo {author} {\bibfnamefont {G.}~\bibnamefont
  {Li}}, \bibinfo {author} {\bibfnamefont {S.}~\bibnamefont {Yin}}, \bibinfo
  {author} {\bibfnamefont {J.}~\bibnamefont {Chen}}, \bibinfo {author}
  {\bibfnamefont {C.}~\bibnamefont {Tan}}, \bibinfo {author} {\bibfnamefont
  {X.}~\bibnamefont {Kan}}, \emph {et~al.},\ }\bibfield  {title} {\bibinfo
  {title} {In-plane anisotropic 2d crps$_4$ for promising
  polarization-sensitive photodetection},\ }\href@noop {} {\bibfield  {journal}
  {\bibinfo  {journal} {Applied Physics Letters}\ }\textbf {\bibinfo {volume}
  {119}},\ \bibinfo {pages} {171102} (\bibinfo {year} {2021})}\BibitemShut
  {NoStop}%
\bibitem [{\citenamefont {Shin}\ \emph {et~al.}(2021)\citenamefont {Shin},
  \citenamefont {Lee}, \citenamefont {Yoon}, \citenamefont {Kim}, \citenamefont
  {Park}, \citenamefont {Lee},\ and\ \citenamefont {Park}}]{shin:21}%
  \BibitemOpen
  \bibfield  {author} {\bibinfo {author} {\bibfnamefont {M.}~\bibnamefont
  {Shin}}, \bibinfo {author} {\bibfnamefont {M.~J.}\ \bibnamefont {Lee}},
  \bibinfo {author} {\bibfnamefont {C.}~\bibnamefont {Yoon}}, \bibinfo {author}
  {\bibfnamefont {S.}~\bibnamefont {Kim}}, \bibinfo {author} {\bibfnamefont
  {B.~H.}\ \bibnamefont {Park}}, \bibinfo {author} {\bibfnamefont
  {S.}~\bibnamefont {Lee}},\ and\ \bibinfo {author} {\bibfnamefont {J.-G.}\
  \bibnamefont {Park}},\ }\bibfield  {title} {\bibinfo {title} {Charge-trapping
  memory device based on a heterostructure of mos$_2$ and crps$_4$},\
  }\href@noop {} {\bibfield  {journal} {\bibinfo  {journal} {Journal of the
  Korean Physical Society}\ }\textbf {\bibinfo {volume} {78}},\ \bibinfo
  {pages} {816} (\bibinfo {year} {2021})}\BibitemShut {NoStop}%
\bibitem [{\citenamefont {Riesner}\ \emph {et~al.}(2022)\citenamefont
  {Riesner}, \citenamefont {Fainblat}, \citenamefont {Budniak}, \citenamefont
  {Amouyal}, \citenamefont {Lifshitz},\ and\ \citenamefont {Bacher}}]{ries:22}%
  \BibitemOpen
  \bibfield  {author} {\bibinfo {author} {\bibfnamefont {M.}~\bibnamefont
  {Riesner}}, \bibinfo {author} {\bibfnamefont {R.}~\bibnamefont {Fainblat}},
  \bibinfo {author} {\bibfnamefont {A.~K.}\ \bibnamefont {Budniak}}, \bibinfo
  {author} {\bibfnamefont {Y.}~\bibnamefont {Amouyal}}, \bibinfo {author}
  {\bibfnamefont {E.}~\bibnamefont {Lifshitz}},\ and\ \bibinfo {author}
  {\bibfnamefont {G.}~\bibnamefont {Bacher}},\ }\bibfield  {title} {\bibinfo
  {title} {Temperature dependence of fano resonances in crps$_4$},\ }\href@noop
  {} {\bibfield  {journal} {\bibinfo  {journal} {The Journal of Chemical
  Physics}\ }\textbf {\bibinfo {volume} {156}},\ \bibinfo {pages} {054707}
  (\bibinfo {year} {2022})}\BibitemShut {NoStop}%
\bibitem [{\citenamefont {Kim}\ \emph {et~al.}(2022)\citenamefont {Kim},
  \citenamefont {Yoon}, \citenamefont {Ahn}, \citenamefont {Jin}, \citenamefont
  {Kim}, \citenamefont {Jo}, \citenamefont {Lee}, \citenamefont {Kim},\ and\
  \citenamefont {Ryu}}]{kim:22}%
  \BibitemOpen
  \bibfield  {author} {\bibinfo {author} {\bibfnamefont {S.}~\bibnamefont
  {Kim}}, \bibinfo {author} {\bibfnamefont {S.}~\bibnamefont {Yoon}}, \bibinfo
  {author} {\bibfnamefont {H.}~\bibnamefont {Ahn}}, \bibinfo {author}
  {\bibfnamefont {G.}~\bibnamefont {Jin}}, \bibinfo {author} {\bibfnamefont
  {H.}~\bibnamefont {Kim}}, \bibinfo {author} {\bibfnamefont {M.-H.}\
  \bibnamefont {Jo}}, \bibinfo {author} {\bibfnamefont {C.}~\bibnamefont
  {Lee}}, \bibinfo {author} {\bibfnamefont {J.}~\bibnamefont {Kim}},\ and\
  \bibinfo {author} {\bibfnamefont {S.}~\bibnamefont {Ryu}},\ }\bibfield
  {title} {\bibinfo {title} {Photoluminescence path bifurcations by spin flip
  in two-dimensional crps$_4$},\ }\href@noop {} {\bibfield  {journal} {\bibinfo
   {journal} {ACS Nano}\ } (\bibinfo {year} {2022})}\BibitemShut {NoStop}%
\bibitem [{\citenamefont {Xu}\ \emph {et~al.}(2022)\citenamefont {Xu},
  \citenamefont {Liu}, \citenamefont {Li}, \citenamefont {Wu}, \citenamefont
  {Zhang}, \citenamefont {Wang}, \citenamefont {Huang},\ and\ \citenamefont
  {Zhang}}]{xu:22}%
  \BibitemOpen
  \bibfield  {author} {\bibinfo {author} {\bibfnamefont {G.}~\bibnamefont
  {Xu}}, \bibinfo {author} {\bibfnamefont {D.}~\bibnamefont {Liu}}, \bibinfo
  {author} {\bibfnamefont {S.}~\bibnamefont {Li}}, \bibinfo {author}
  {\bibfnamefont {Y.}~\bibnamefont {Wu}}, \bibinfo {author} {\bibfnamefont
  {Z.}~\bibnamefont {Zhang}}, \bibinfo {author} {\bibfnamefont
  {S.}~\bibnamefont {Wang}}, \bibinfo {author} {\bibfnamefont {Z.}~\bibnamefont
  {Huang}},\ and\ \bibinfo {author} {\bibfnamefont {Y.}~\bibnamefont {Zhang}},\
  }\bibfield  {title} {\bibinfo {title} {Binary-ternary transition metal
  chalcogenides interlayer coupling in van der waals type-ii heterostructure
  for visible-infrared photodetector with efficient suppression dark
  currents},\ }\href@noop {} {\bibfield  {journal} {\bibinfo  {journal} {Nano
  Research}\ }\textbf {\bibinfo {volume} {15}},\ \bibinfo {pages} {2689}
  (\bibinfo {year} {2022})}\BibitemShut {NoStop}%
\bibitem [{\citenamefont {Zhuang}\ and\ \citenamefont {Zhou}(2016)}]{zhua:16}%
  \BibitemOpen
  \bibfield  {author} {\bibinfo {author} {\bibfnamefont {H.}~\bibnamefont
  {Zhuang}}\ and\ \bibinfo {author} {\bibfnamefont {J.}~\bibnamefont {Zhou}},\
  }\bibfield  {title} {\bibinfo {title} {{Density functional theory study of
  bulk and single-layer magnetic semiconductor CrPS$_{4}$}},\ }\href@noop {}
  {\bibfield  {journal} {\bibinfo  {journal} {Physical Review B}\ }\textbf
  {\bibinfo {volume} {94}},\ \bibinfo {pages} {195307} (\bibinfo {year}
  {2016})}\BibitemShut {NoStop}%
\bibitem [{\citenamefont {Joe}\ \emph {et~al.}(2017)\citenamefont {Joe},
  \citenamefont {Lee}, \citenamefont {Alyoruk}, \citenamefont {Lee},
  \citenamefont {Kim}, \citenamefont {Lee},\ and\ \citenamefont
  {Lee}}]{joe:17}%
  \BibitemOpen
  \bibfield  {author} {\bibinfo {author} {\bibfnamefont {M.}~\bibnamefont
  {Joe}}, \bibinfo {author} {\bibfnamefont {H.}~\bibnamefont {Lee}}, \bibinfo
  {author} {\bibfnamefont {M.}~\bibnamefont {Alyoruk}}, \bibinfo {author}
  {\bibfnamefont {J.}~\bibnamefont {Lee}}, \bibinfo {author} {\bibfnamefont
  {S.}~\bibnamefont {Kim}}, \bibinfo {author} {\bibfnamefont {C.}~\bibnamefont
  {Lee}},\ and\ \bibinfo {author} {\bibfnamefont {J.}~\bibnamefont {Lee}},\
  }\bibfield  {title} {\bibinfo {title} {{A comprehensive study of
  piezomagnetic response in CrPS$_{4}$ monolayer: mechanical, electronic
  properties and magnetic ordering under strains}},\ }\href@noop {} {\bibfield
  {journal} {\bibinfo  {journal} {JoP: Condensed Matter}\ }\textbf {\bibinfo
  {volume} {29}},\ \bibinfo {pages} {405801} (\bibinfo {year}
  {2017})}\BibitemShut {NoStop}%
\bibitem [{\citenamefont {Chen}\ \emph {et~al.}(2020)\citenamefont {Chen},
  \citenamefont {Ding}, \citenamefont {Wang}, \citenamefont {Xu},\ and\
  \citenamefont {Wang}}]{chen:20}%
  \BibitemOpen
  \bibfield  {author} {\bibinfo {author} {\bibfnamefont {Q.}~\bibnamefont
  {Chen}}, \bibinfo {author} {\bibfnamefont {Q.}~\bibnamefont {Ding}}, \bibinfo
  {author} {\bibfnamefont {Y.}~\bibnamefont {Wang}}, \bibinfo {author}
  {\bibfnamefont {Y.}~\bibnamefont {Xu}},\ and\ \bibinfo {author}
  {\bibfnamefont {J.}~\bibnamefont {Wang}},\ }\bibfield  {title} {\bibinfo
  {title} {{Electronic and Magnetic Properties of a Two-Dimensional Transition
  Metal Phosphorous Chalcogenide TMPS$_{4}$}},\ }\href@noop {} {\bibfield
  {journal} {\bibinfo  {journal} {J. Phys. Chem. C}\ }\textbf {\bibinfo
  {volume} {124}},\ \bibinfo {pages} {12075} (\bibinfo {year}
  {2020})}\BibitemShut {NoStop}%
\bibitem [{\citenamefont {Deng}\ \emph {et~al.}(2021)\citenamefont {Deng},
  \citenamefont {Guo}, \citenamefont {Hosono}, \citenamefont {Ying},\ and\
  \citenamefont {Chen}}]{deng:21}%
  \BibitemOpen
  \bibfield  {author} {\bibinfo {author} {\bibfnamefont {J.}~\bibnamefont
  {Deng}}, \bibinfo {author} {\bibfnamefont {J.}~\bibnamefont {Guo}}, \bibinfo
  {author} {\bibfnamefont {H.}~\bibnamefont {Hosono}}, \bibinfo {author}
  {\bibfnamefont {T.}~\bibnamefont {Ying}},\ and\ \bibinfo {author}
  {\bibfnamefont {X.}~\bibnamefont {Chen}},\ }\bibfield  {title} {\bibinfo
  {title} {Two-dimensional bipolar ferromagnetic semiconductors from layered
  antiferromagnets},\ }\href@noop {} {\bibfield  {journal} {\bibinfo  {journal}
  {Physical Review Materials}\ }\textbf {\bibinfo {volume} {5}},\ \bibinfo
  {pages} {034005} (\bibinfo {year} {2021})}\BibitemShut {NoStop}%
\bibitem [{\citenamefont {Yang}\ \emph {et~al.}(2021)\citenamefont {Yang},
  \citenamefont {Fang}, \citenamefont {Peng}, \citenamefont {Liu},
  \citenamefont {Wu}, \citenamefont {Quhe}, \citenamefont {Ding}, \citenamefont
  {Yang}, \citenamefont {Ma}, \citenamefont {Shi} \emph {et~al.}}]{yang:21}%
  \BibitemOpen
  \bibfield  {author} {\bibinfo {author} {\bibfnamefont {J.}~\bibnamefont
  {Yang}}, \bibinfo {author} {\bibfnamefont {S.}~\bibnamefont {Fang}}, \bibinfo
  {author} {\bibfnamefont {Y.}~\bibnamefont {Peng}}, \bibinfo {author}
  {\bibfnamefont {S.}~\bibnamefont {Liu}}, \bibinfo {author} {\bibfnamefont
  {B.}~\bibnamefont {Wu}}, \bibinfo {author} {\bibfnamefont {R.}~\bibnamefont
  {Quhe}}, \bibinfo {author} {\bibfnamefont {S.}~\bibnamefont {Ding}}, \bibinfo
  {author} {\bibfnamefont {C.}~\bibnamefont {Yang}}, \bibinfo {author}
  {\bibfnamefont {J.}~\bibnamefont {Ma}}, \bibinfo {author} {\bibfnamefont
  {B.}~\bibnamefont {Shi}}, \emph {et~al.},\ }\bibfield  {title} {\bibinfo
  {title} {Layer-dependent giant magnetoresistance in two-dimensional crps$_4$
  magnetic tunnel junctions},\ }\href@noop {} {\bibfield  {journal} {\bibinfo
  {journal} {Physical Review Applied}\ }\textbf {\bibinfo {volume} {16}},\
  \bibinfo {pages} {024011} (\bibinfo {year} {2021})}\BibitemShut {NoStop}%
\bibitem [{\citenamefont {Louisy}\ \emph {et~al.}(1978)\citenamefont {Louisy},
  \citenamefont {Ouvrard}, \citenamefont {Schleich},\ and\ \citenamefont
  {Brec}}]{loui:78}%
  \BibitemOpen
  \bibfield  {author} {\bibinfo {author} {\bibfnamefont {A.}~\bibnamefont
  {Louisy}}, \bibinfo {author} {\bibfnamefont {G.}~\bibnamefont {Ouvrard}},
  \bibinfo {author} {\bibfnamefont {D.}~\bibnamefont {Schleich}},\ and\
  \bibinfo {author} {\bibfnamefont {R.}~\bibnamefont {Brec}},\ }\bibfield
  {title} {\bibinfo {title} {{Physical properties and lithium intercalates of
  CrPS$_{4}$}},\ }\href@noop {} {\bibfield  {journal} {\bibinfo  {journal}
  {Solid State Communications}\ }\textbf {\bibinfo {volume} {28}},\ \bibinfo
  {pages} {61} (\bibinfo {year} {1978})}\BibitemShut {NoStop}%
\bibitem [{\citenamefont {Lee}\ \emph {et~al.}(2018)\citenamefont {Lee},
  \citenamefont {Lee}, \citenamefont {Lee}, \citenamefont {Balamurugan},
  \citenamefont {Yoon}, \citenamefont {Jang}, \citenamefont {Kim},
  \citenamefont {Kwon}, \citenamefont {Kim}, \citenamefont {Ahn}, \citenamefont
  {Kim}, \citenamefont {Park},\ and\ \citenamefont {Park}}]{lee:18}%
  \BibitemOpen
  \bibfield  {author} {\bibinfo {author} {\bibfnamefont {M.}~\bibnamefont
  {Lee}}, \bibinfo {author} {\bibfnamefont {S.}~\bibnamefont {Lee}}, \bibinfo
  {author} {\bibfnamefont {S.}~\bibnamefont {Lee}}, \bibinfo {author}
  {\bibfnamefont {K.}~\bibnamefont {Balamurugan}}, \bibinfo {author}
  {\bibfnamefont {C.}~\bibnamefont {Yoon}}, \bibinfo {author} {\bibfnamefont
  {J.}~\bibnamefont {Jang}}, \bibinfo {author} {\bibfnamefont {S.}~\bibnamefont
  {Kim}}, \bibinfo {author} {\bibfnamefont {D.}~\bibnamefont {Kwon}}, \bibinfo
  {author} {\bibfnamefont {M.}~\bibnamefont {Kim}}, \bibinfo {author}
  {\bibfnamefont {J.}~\bibnamefont {Ahn}}, \bibinfo {author} {\bibfnamefont
  {D.}~\bibnamefont {Kim}}, \bibinfo {author} {\bibfnamefont {J.}~\bibnamefont
  {Park}},\ and\ \bibinfo {author} {\bibfnamefont {B.}~\bibnamefont {Park}},\
  }\bibfield  {title} {\bibinfo {title} {{Synaptic devices based on
  two-dimensional layered single-crystal chromium thiophosphate
  (CrPS$_{4}$)}},\ }\href@noop {} {\bibfield  {journal} {\bibinfo  {journal}
  {NPG Asia Materials}\ }\textbf {\bibinfo {volume} {10}},\ \bibinfo {pages}
  {23} (\bibinfo {year} {2018})}\BibitemShut {NoStop}%
\bibitem [{\citenamefont {Kresse}\ and\ \citenamefont
  {Joubert}(1999)}]{kres:99}%
  \BibitemOpen
  \bibfield  {author} {\bibinfo {author} {\bibfnamefont {G.}~\bibnamefont
  {Kresse}}\ and\ \bibinfo {author} {\bibfnamefont {D.}~\bibnamefont
  {Joubert}},\ }\bibfield  {title} {\bibinfo {title} {{From ultrasoft
  pseudopotentials to the projector augmented-wave method}},\ }\href@noop {}
  {\bibfield  {journal} {\bibinfo  {journal} {Physical Review B}\ }\textbf
  {\bibinfo {volume} {59}},\ \bibinfo {pages} {1758} (\bibinfo {year}
  {1999})}\BibitemShut {NoStop}%
\bibitem [{\citenamefont {Kresse}\ and\ \citenamefont
  {Hafner}(1993)}]{kres:93}%
  \BibitemOpen
  \bibfield  {author} {\bibinfo {author} {\bibfnamefont {G.}~\bibnamefont
  {Kresse}}\ and\ \bibinfo {author} {\bibfnamefont {J.}~\bibnamefont
  {Hafner}},\ }\bibfield  {title} {\bibinfo {title} {{$Ab$ $initio$ molecular
  dynamics for open-shell transition metals}},\ }\href@noop {} {\bibfield
  {journal} {\bibinfo  {journal} {Physical Review B}\ }\textbf {\bibinfo
  {volume} {48}},\ \bibinfo {pages} {13115} (\bibinfo {year}
  {1993})}\BibitemShut {NoStop}%
\bibitem [{\citenamefont {Kresse}\ and\ \citenamefont
  {Furthmuller}(1996)}]{kres:96}%
  \BibitemOpen
  \bibfield  {author} {\bibinfo {author} {\bibfnamefont {G.}~\bibnamefont
  {Kresse}}\ and\ \bibinfo {author} {\bibfnamefont {J.}~\bibnamefont
  {Furthmuller}},\ }\bibfield  {title} {\bibinfo {title} {{Efficient iterative
  schemes for $ab$ $initio$ total-energy calculations using a plane-wave basis
  set}},\ }\href@noop {} {\bibfield  {journal} {\bibinfo  {journal} {Physical
  Review B}\ }\textbf {\bibinfo {volume} {54}},\ \bibinfo {pages} {11169}
  (\bibinfo {year} {1996})}\BibitemShut {NoStop}%
\bibitem [{\citenamefont {Sun}\ \emph {et~al.}(2015)\citenamefont {Sun},
  \citenamefont {Ruzsinszky},\ and\ \citenamefont {Perdew}}]{sunj:15}%
  \BibitemOpen
  \bibfield  {author} {\bibinfo {author} {\bibfnamefont {J.}~\bibnamefont
  {Sun}}, \bibinfo {author} {\bibfnamefont {A.}~\bibnamefont {Ruzsinszky}},\
  and\ \bibinfo {author} {\bibfnamefont {J.}~\bibnamefont {Perdew}},\
  }\bibfield  {title} {\bibinfo {title} {{Strongly Constrained and
  Appropriately Normed Semilocal Density Functional}},\ }\href@noop {}
  {\bibfield  {journal} {\bibinfo  {journal} {Physical Review Letters}\
  }\textbf {\bibinfo {volume} {115}},\ \bibinfo {pages} {036402} (\bibinfo
  {year} {2015})}\BibitemShut {NoStop}%
\bibitem [{\citenamefont {Diehl}\ and\ \citenamefont
  {Carpentier}(1976)}]{dieh:76}%
  \BibitemOpen
  \bibfield  {author} {\bibinfo {author} {\bibfnamefont {R.}~\bibnamefont
  {Diehl}}\ and\ \bibinfo {author} {\bibfnamefont {C.}~\bibnamefont
  {Carpentier}},\ }\bibfield  {title} {\bibinfo {title} {{The Crystal Structure
  of Chromium Thiophosphate, CrPS$_{4}$}},\ }\href@noop {} {\bibfield
  {journal} {\bibinfo  {journal} {Acta Crystallographica Section B}\ }\textbf
  {\bibinfo {volume} {B33}},\ \bibinfo {pages} {1399} (\bibinfo {year}
  {1976})}\BibitemShut {NoStop}%
\bibitem [{\citenamefont {Kohn}\ and\ \citenamefont
  {Sham}(1965)}]{kohn1965self}%
  \BibitemOpen
  \bibfield  {author} {\bibinfo {author} {\bibfnamefont {W.}~\bibnamefont
  {Kohn}}\ and\ \bibinfo {author} {\bibfnamefont {L.~J.}\ \bibnamefont
  {Sham}},\ }\bibfield  {title} {\bibinfo {title} {Self-consistent equations
  including exchange and correlation effects},\ }\href@noop {} {\bibfield
  {journal} {\bibinfo  {journal} {Physical review}\ }\textbf {\bibinfo {volume}
  {140}},\ \bibinfo {pages} {A1133} (\bibinfo {year} {1965})}\BibitemShut
  {NoStop}%
\bibitem [{\citenamefont {Seidl}\ \emph {et~al.}(1996)\citenamefont {Seidl},
  \citenamefont {G{\"o}rling}, \citenamefont {Vogl}, \citenamefont {Majewski},\
  and\ \citenamefont {Levy}}]{seidl1996generalized}%
  \BibitemOpen
  \bibfield  {author} {\bibinfo {author} {\bibfnamefont {A.}~\bibnamefont
  {Seidl}}, \bibinfo {author} {\bibfnamefont {A.}~\bibnamefont {G{\"o}rling}},
  \bibinfo {author} {\bibfnamefont {P.}~\bibnamefont {Vogl}}, \bibinfo {author}
  {\bibfnamefont {J.~A.}\ \bibnamefont {Majewski}},\ and\ \bibinfo {author}
  {\bibfnamefont {M.}~\bibnamefont {Levy}},\ }\bibfield  {title} {\bibinfo
  {title} {Generalized kohn-sham schemes and the band-gap problem},\
  }\href@noop {} {\bibfield  {journal} {\bibinfo  {journal} {Physical Review
  B}\ }\textbf {\bibinfo {volume} {53}},\ \bibinfo {pages} {3764} (\bibinfo
  {year} {1996})}\BibitemShut {NoStop}%
\bibitem [{\citenamefont {Perdew}\ \emph {et~al.}(2017)\citenamefont {Perdew},
  \citenamefont {Yang}, \citenamefont {Burke}, \citenamefont {Yang},
  \citenamefont {Gross}, \citenamefont {Scheffler}, \citenamefont {Scuseria},
  \citenamefont {Henderson}, \citenamefont {Zhang}, \citenamefont {Ruzsinszky}
  \emph {et~al.}}]{perd:17}%
  \BibitemOpen
  \bibfield  {author} {\bibinfo {author} {\bibfnamefont {J.~P.}\ \bibnamefont
  {Perdew}}, \bibinfo {author} {\bibfnamefont {W.}~\bibnamefont {Yang}},
  \bibinfo {author} {\bibfnamefont {K.}~\bibnamefont {Burke}}, \bibinfo
  {author} {\bibfnamefont {Z.}~\bibnamefont {Yang}}, \bibinfo {author}
  {\bibfnamefont {E.~K.}\ \bibnamefont {Gross}}, \bibinfo {author}
  {\bibfnamefont {M.}~\bibnamefont {Scheffler}}, \bibinfo {author}
  {\bibfnamefont {G.~E.}\ \bibnamefont {Scuseria}}, \bibinfo {author}
  {\bibfnamefont {T.~M.}\ \bibnamefont {Henderson}}, \bibinfo {author}
  {\bibfnamefont {I.~Y.}\ \bibnamefont {Zhang}}, \bibinfo {author}
  {\bibfnamefont {A.}~\bibnamefont {Ruzsinszky}}, \emph {et~al.},\ }\bibfield
  {title} {\bibinfo {title} {Understanding band gaps of solids in generalized
  kohn--sham theory},\ }\href@noop {} {\bibfield  {journal} {\bibinfo
  {journal} {Proceedings of the National Academy of Sciences}\ }\textbf
  {\bibinfo {volume} {114}},\ \bibinfo {pages} {2801} (\bibinfo {year}
  {2017})}\BibitemShut {NoStop}%
\bibitem [{\citenamefont {Furness}\ \emph {et~al.}(2018)\citenamefont
  {Furness}, \citenamefont {Zhang}, \citenamefont {Lane}, \citenamefont {Buda},
  \citenamefont {Barbiellini}, \citenamefont {Markiewicz}, \citenamefont
  {Bansil},\ and\ \citenamefont {Sun}}]{furness2018accurate}%
  \BibitemOpen
  \bibfield  {author} {\bibinfo {author} {\bibfnamefont {J.~W.}\ \bibnamefont
  {Furness}}, \bibinfo {author} {\bibfnamefont {Y.}~\bibnamefont {Zhang}},
  \bibinfo {author} {\bibfnamefont {C.}~\bibnamefont {Lane}}, \bibinfo {author}
  {\bibfnamefont {I.~G.}\ \bibnamefont {Buda}}, \bibinfo {author}
  {\bibfnamefont {B.}~\bibnamefont {Barbiellini}}, \bibinfo {author}
  {\bibfnamefont {R.~S.}\ \bibnamefont {Markiewicz}}, \bibinfo {author}
  {\bibfnamefont {A.}~\bibnamefont {Bansil}},\ and\ \bibinfo {author}
  {\bibfnamefont {J.}~\bibnamefont {Sun}},\ }\bibfield  {title} {\bibinfo
  {title} {An accurate first-principles treatment of doping-dependent
  electronic structure of high-temperature cuprate superconductors},\
  }\href@noop {} {\bibfield  {journal} {\bibinfo  {journal} {Communications
  Physics}\ }\textbf {\bibinfo {volume} {1}},\ \bibinfo {pages} {11} (\bibinfo
  {year} {2018})}\BibitemShut {NoStop}%
\bibitem [{\citenamefont {Lane}\ \emph {et~al.}(2018)\citenamefont {Lane},
  \citenamefont {Furness}, \citenamefont {Buda}, \citenamefont {Zhang},
  \citenamefont {Markiewicz}, \citenamefont {Barbiellini}, \citenamefont
  {Sun},\ and\ \citenamefont {Bansil}}]{lane2018antiferromagnetic}%
  \BibitemOpen
  \bibfield  {author} {\bibinfo {author} {\bibfnamefont {C.}~\bibnamefont
  {Lane}}, \bibinfo {author} {\bibfnamefont {J.~W.}\ \bibnamefont {Furness}},
  \bibinfo {author} {\bibfnamefont {I.~G.}\ \bibnamefont {Buda}}, \bibinfo
  {author} {\bibfnamefont {Y.}~\bibnamefont {Zhang}}, \bibinfo {author}
  {\bibfnamefont {R.~S.}\ \bibnamefont {Markiewicz}}, \bibinfo {author}
  {\bibfnamefont {B.}~\bibnamefont {Barbiellini}}, \bibinfo {author}
  {\bibfnamefont {J.}~\bibnamefont {Sun}},\ and\ \bibinfo {author}
  {\bibfnamefont {A.}~\bibnamefont {Bansil}},\ }\bibfield  {title} {\bibinfo
  {title} {Antiferromagnetic ground state of la 2 cuo 4: A parameter-free ab
  initio description},\ }\href@noop {} {\bibfield  {journal} {\bibinfo
  {journal} {Physical Review B}\ }\textbf {\bibinfo {volume} {98}},\ \bibinfo
  {pages} {125140} (\bibinfo {year} {2018})}\BibitemShut {NoStop}%
\bibitem [{\citenamefont {Pokharel}\ \emph {et~al.}(2022)\citenamefont
  {Pokharel}, \citenamefont {Lane}, \citenamefont {Furness}, \citenamefont
  {Zhang}, \citenamefont {Ning}, \citenamefont {Barbiellini}, \citenamefont
  {Markiewicz}, \citenamefont {Zhang}, \citenamefont {Bansil},\ and\
  \citenamefont {Sun}}]{pokharel2022sensitivity}%
  \BibitemOpen
  \bibfield  {author} {\bibinfo {author} {\bibfnamefont {K.}~\bibnamefont
  {Pokharel}}, \bibinfo {author} {\bibfnamefont {C.}~\bibnamefont {Lane}},
  \bibinfo {author} {\bibfnamefont {J.~W.}\ \bibnamefont {Furness}}, \bibinfo
  {author} {\bibfnamefont {R.}~\bibnamefont {Zhang}}, \bibinfo {author}
  {\bibfnamefont {J.}~\bibnamefont {Ning}}, \bibinfo {author} {\bibfnamefont
  {B.}~\bibnamefont {Barbiellini}}, \bibinfo {author} {\bibfnamefont {R.~S.}\
  \bibnamefont {Markiewicz}}, \bibinfo {author} {\bibfnamefont
  {Y.}~\bibnamefont {Zhang}}, \bibinfo {author} {\bibfnamefont
  {A.}~\bibnamefont {Bansil}},\ and\ \bibinfo {author} {\bibfnamefont
  {J.}~\bibnamefont {Sun}},\ }\bibfield  {title} {\bibinfo {title} {Sensitivity
  of the electronic and magnetic structures of cuprate superconductors to
  density functional approximations},\ }\href@noop {} {\bibfield  {journal}
  {\bibinfo  {journal} {npj Computational Materials}\ }\textbf {\bibinfo
  {volume} {8}},\ \bibinfo {pages} {31} (\bibinfo {year} {2022})}\BibitemShut
  {NoStop}%
\bibitem [{\citenamefont {Zhang}\ \emph {et~al.}(2020)\citenamefont {Zhang},
  \citenamefont {Lane}, \citenamefont {Furness}, \citenamefont {Barbiellini},
  \citenamefont {Perdew}, \citenamefont {Markiewicz}, \citenamefont {Bansil},\
  and\ \citenamefont {Sun}}]{zhang2020competing}%
  \BibitemOpen
  \bibfield  {author} {\bibinfo {author} {\bibfnamefont {Y.}~\bibnamefont
  {Zhang}}, \bibinfo {author} {\bibfnamefont {C.}~\bibnamefont {Lane}},
  \bibinfo {author} {\bibfnamefont {J.~W.}\ \bibnamefont {Furness}}, \bibinfo
  {author} {\bibfnamefont {B.}~\bibnamefont {Barbiellini}}, \bibinfo {author}
  {\bibfnamefont {J.~P.}\ \bibnamefont {Perdew}}, \bibinfo {author}
  {\bibfnamefont {R.~S.}\ \bibnamefont {Markiewicz}}, \bibinfo {author}
  {\bibfnamefont {A.}~\bibnamefont {Bansil}},\ and\ \bibinfo {author}
  {\bibfnamefont {J.}~\bibnamefont {Sun}},\ }\bibfield  {title} {\bibinfo
  {title} {Competing stripe and magnetic phases in the cuprates from first
  principles},\ }\href@noop {} {\bibfield  {journal} {\bibinfo  {journal}
  {Proceedings of the National Academy of Sciences}\ }\textbf {\bibinfo
  {volume} {117}},\ \bibinfo {pages} {68} (\bibinfo {year} {2020})}\BibitemShut
  {NoStop}%
\bibitem [{\citenamefont {Lane}\ \emph {et~al.}(2020)\citenamefont {Lane},
  \citenamefont {Zhang}, \citenamefont {Furness}, \citenamefont {Markiewicz},
  \citenamefont {Barbiellini}, \citenamefont {Sun},\ and\ \citenamefont
  {Bansil}}]{lane2020first}%
  \BibitemOpen
  \bibfield  {author} {\bibinfo {author} {\bibfnamefont {C.}~\bibnamefont
  {Lane}}, \bibinfo {author} {\bibfnamefont {Y.}~\bibnamefont {Zhang}},
  \bibinfo {author} {\bibfnamefont {J.~W.}\ \bibnamefont {Furness}}, \bibinfo
  {author} {\bibfnamefont {R.~S.}\ \bibnamefont {Markiewicz}}, \bibinfo
  {author} {\bibfnamefont {B.}~\bibnamefont {Barbiellini}}, \bibinfo {author}
  {\bibfnamefont {J.}~\bibnamefont {Sun}},\ and\ \bibinfo {author}
  {\bibfnamefont {A.}~\bibnamefont {Bansil}},\ }\bibfield  {title} {\bibinfo
  {title} {First-principles calculation of spin and orbital contributions to
  magnetically ordered moments in sr 2 iro 4},\ }\href@noop {} {\bibfield
  {journal} {\bibinfo  {journal} {Physical Review B}\ }\textbf {\bibinfo
  {volume} {101}},\ \bibinfo {pages} {155110} (\bibinfo {year}
  {2020})}\BibitemShut {NoStop}%
\bibitem [{\citenamefont {Lane}\ and\ \citenamefont
  {Zhu}(2020{\natexlab{a}})}]{lane2020thickness}%
  \BibitemOpen
  \bibfield  {author} {\bibinfo {author} {\bibfnamefont {C.}~\bibnamefont
  {Lane}}\ and\ \bibinfo {author} {\bibfnamefont {J.-X.}\ \bibnamefont {Zhu}},\
  }\bibfield  {title} {\bibinfo {title} {Thickness dependence of electronic
  structure and optical properties of a correlated van der waals
  antiferromagnetic nips$_3$ thin film},\ }\href@noop {} {\bibfield  {journal}
  {\bibinfo  {journal} {Physical Review B}\ }\textbf {\bibinfo {volume}
  {102}},\ \bibinfo {pages} {075124} (\bibinfo {year}
  {2020}{\natexlab{a}})}\BibitemShut {NoStop}%
\bibitem [{\citenamefont {Lane}\ and\ \citenamefont
  {Zhu}(2020{\natexlab{b}})}]{lane2020landscape}%
  \BibitemOpen
  \bibfield  {author} {\bibinfo {author} {\bibfnamefont {C.}~\bibnamefont
  {Lane}}\ and\ \bibinfo {author} {\bibfnamefont {J.-X.}\ \bibnamefont {Zhu}},\
  }\bibfield  {title} {\bibinfo {title} {Landscape of coexisting excitonic
  states in the insulating single-layer cuprates and nickelates},\ }\href@noop
  {} {\bibfield  {journal} {\bibinfo  {journal} {Physical Review B}\ }\textbf
  {\bibinfo {volume} {101}},\ \bibinfo {pages} {155135} (\bibinfo {year}
  {2020}{\natexlab{b}})}\BibitemShut {NoStop}%
\bibitem [{\citenamefont {Lane}\ and\ \citenamefont {Zhu}(2022)}]{lane2022ab}%
  \BibitemOpen
  \bibfield  {author} {\bibinfo {author} {\bibfnamefont {C.}~\bibnamefont
  {Lane}}\ and\ \bibinfo {author} {\bibfnamefont {J.-X.}\ \bibnamefont {Zhu}},\
  }\bibfield  {title} {\bibinfo {title} {An ab initio study of electron-hole
  pairs in a correlated van der waals antiferromagnet: Nips $ \_3$},\
  }\href@noop {} {\bibfield  {journal} {\bibinfo  {journal} {arXiv preprint
  arXiv:2209.13051}\ } (\bibinfo {year} {2022})}\BibitemShut {NoStop}%
\bibitem [{\citenamefont {Zaanen}\ \emph {et~al.}(1985)\citenamefont {Zaanen},
  \citenamefont {Sawatzky},\ and\ \citenamefont {Allen}}]{zaan:85}%
  \BibitemOpen
  \bibfield  {author} {\bibinfo {author} {\bibfnamefont {J.}~\bibnamefont
  {Zaanen}}, \bibinfo {author} {\bibfnamefont {G.}~\bibnamefont {Sawatzky}},\
  and\ \bibinfo {author} {\bibfnamefont {J.}~\bibnamefont {Allen}},\ }\bibfield
   {title} {\bibinfo {title} {Band gaps and electronic structure of
  transition-metal compounds},\ }\href@noop {} {\bibfield  {journal} {\bibinfo
  {journal} {Physical Review Letters}\ }\textbf {\bibinfo {volume} {55}},\
  \bibinfo {pages} {418} (\bibinfo {year} {1985})}\BibitemShut {NoStop}%
\bibitem [{\citenamefont {Erskine}\ and\ \citenamefont
  {Stern}(1975)}]{ersk:75}%
  \BibitemOpen
  \bibfield  {author} {\bibinfo {author} {\bibfnamefont {J.~L.}\ \bibnamefont
  {Erskine}}\ and\ \bibinfo {author} {\bibfnamefont {E.}~\bibnamefont
  {Stern}},\ }\bibfield  {title} {\bibinfo {title} {Calculation of the m 23
  magneto-optical absorption spectrum of ferromagnetic nickel},\ }\href@noop {}
  {\bibfield  {journal} {\bibinfo  {journal} {Physical Review B}\ }\textbf
  {\bibinfo {volume} {12}},\ \bibinfo {pages} {5016} (\bibinfo {year}
  {1975})}\BibitemShut {NoStop}%
\bibitem [{\citenamefont {Xia}\ \emph {et~al.}(2006)\citenamefont {Xia},
  \citenamefont {Maeno}, \citenamefont {Beyersdorf}, \citenamefont {Fejer},\
  and\ \citenamefont {Kapitulnik}}]{xia:06}%
  \BibitemOpen
  \bibfield  {author} {\bibinfo {author} {\bibfnamefont {J.}~\bibnamefont
  {Xia}}, \bibinfo {author} {\bibfnamefont {Y.}~\bibnamefont {Maeno}}, \bibinfo
  {author} {\bibfnamefont {P.~T.}\ \bibnamefont {Beyersdorf}}, \bibinfo
  {author} {\bibfnamefont {M.}~\bibnamefont {Fejer}},\ and\ \bibinfo {author}
  {\bibfnamefont {A.}~\bibnamefont {Kapitulnik}},\ }\bibfield  {title}
  {\bibinfo {title} {High resolution polar kerr effect measurements of sr$_2$
  ruo$_4$: Evidence for broken time-reversal symmetry in the superconducting
  state},\ }\href@noop {} {\bibfield  {journal} {\bibinfo  {journal} {Physical
  Review Letters}\ }\textbf {\bibinfo {volume} {97}},\ \bibinfo {pages}
  {167002} (\bibinfo {year} {2006})}\BibitemShut {NoStop}%
\bibitem [{\citenamefont {Sangalli}\ \emph {et~al.}(2012)\citenamefont
  {Sangalli}, \citenamefont {Marini},\ and\ \citenamefont
  {Debernardi}}]{sang:12}%
  \BibitemOpen
  \bibfield  {author} {\bibinfo {author} {\bibfnamefont {D.}~\bibnamefont
  {Sangalli}}, \bibinfo {author} {\bibfnamefont {A.}~\bibnamefont {Marini}},\
  and\ \bibinfo {author} {\bibfnamefont {A.}~\bibnamefont {Debernardi}},\
  }\bibfield  {title} {\bibinfo {title} {Pseudopotential-based first-principles
  approach to the magneto-optical kerr effect: From metals to the inclusion of
  local fields and excitonic effects},\ }\href@noop {} {\bibfield  {journal}
  {\bibinfo  {journal} {Physical Review B}\ }\textbf {\bibinfo {volume} {86}},\
  \bibinfo {pages} {125139} (\bibinfo {year} {2012})}\BibitemShut {NoStop}%
\end{thebibliography}%

\end{document}